\documentclass[useAMS,usenatbib,usefloat]{mnras}
\usepackage{float}
\usepackage{amsmath}
\usepackage{amssymb}
\usepackage{rotating}
\usepackage{tabularx}
\usepackage{gensymb}

\include{journaldefs}
\usepackage[font=footnotesize,labelfont=bf]{caption}

\begin{document}

\title[\bf O stars around Westerlund 2]{\bf Massive stars in the hinterland of the young cluster, Westerlund 2}
\author[J. E. Drew et al]{
{\parbox{\textwidth}{J. E. Drew,$^{1}$\thanks{E-mail: j.drew@herts.ac.uk}, A. Herrero$^{2,3}$, M. Mohr-Smith$^1$, M. Mongui\'o$^1$, N. J. Wright$^4$,\\T. Kupfer$^{5,6}$, R. Napiwotzki$^1$}
}\\ \\
$^{1}$School of Physics, Astronomy \& Mathematics, University of Hertfordshire, Hatfield AL10 9AB, UK \\
$^2$Instituto de Astrofis\'ica de Canarias, 38200, La Laguna, Tenerife, Spain\\
$^3$Departamento de Astrof\'isica, Universidad de La Laguna, 38205, La Laguna, Tenerife, Spain\\
$^4$Astrophysics Group, Keele University, Keele, ST5 5BG, UK\\
$^5$Kavli Institute for Theoretical Physics, University of California, Santa Barbara, CA 93106, USA\\
$^6$Cahill Center for Astronomy \& Astrophysics, California Institute of Technology, Pasadena, CA 91125, USA\\
}

\maketitle

\begin{abstract}

An unsettled question concerning the formation and distribution of massive stars is whether they must be born in massive clusters and, if found in less dense environments, whether they must have migrated there.  With the advent of wide-area digital photometric surveys, it is now possible to identify massive stars away from prominent Galactic clusters without bias.  In this study we consider 40 candidate OB stars found in the field around the young massive cluster, Westerlund 2, by Mohr-Smith et al (2017): these are located inside a box of 1.5$\times$1.5 square degrees and are selected on the basis of their extinctions and $K$ magnitudes.  We present VLT/X-shooter spectra of two of the hottest O stars, respectively 11 and 22 arcmin from the centre of Westerlund 2.  They are confirmed as O4V stars, with stellar masses likely to be in excess of 40~M$_{\odot}$. Their radial velocities relative to the non-binary reference object, MSP 182, in Westerlund 2 are $-$29.4$\pm$1.7 and $-$14.4$\pm$2.2 km s$^{-1}$, respectively.  Using Gaia DR2 proper motions we find that between 8 and 11 early O/WR stars in the studied region (including the two VLT targets, plus WR 20c and WR 20aa) could have been ejected from Westerlund 2 in the last one million years.  This represents an efficiency of massive-star ejection of up to $\sim$ 25\%.  On sky, the positions of these stars and their proper motions show a near N--S alignment.  We discuss the possibility that these results are a consequence of prior sub-cluster merging combining with dynamical ejection.
%A potential background OB association with $A_0 \sim 8$ is picked out 
%around Galactic coordinates $\ell = 283.8^{\circ}$, $b = -0.9^{\circ}$. 
%Conclusion - need 2 or more lines to avoid pdfendlink, pdfstartlink problem

\end{abstract}


%\clearpage

\section{Introduction}

Massive OB stars are critically important objects in shaping the evolution of galactic environments: it has long been recognised that the UV radiation, supernovae and winds they produce play leading roles in shaping the interstellar medium (ISM).  But because of their relative rarity at formation and their short lives, their properties have proved difficult to nail down. Indeed, their mode or modes of formation are still unclear \citep{Tan2014}.  The debate is fuelled by questions concerning the impact of binarity on both evolution and dynamics \citep{Sana2012} and the implications of the stark environmental contrast between dense young clusters and looser OB associations as formation sites \citep[e.g.][]{Wright2014}.  To add to this, there remains a lack of clarity over the possibility that some fraction of massive stars could even form in relative isolation \citep{deWit2005,Bressert2012,Gvaramadze2012}.  In this last case, the vagaries of random sampling of the initial mass function (IMF) may be relevant \citep{Parker2007}.

It is a common perception that the centre of more massive clusters is the preferred birthplace for most O stars ($T_{eff} > 30 kK$ on the main sequence, our working definition of an O star).  In this context the phenomenon of runaway O stars \citep{Blaauw1961} slots into place as the explanation for massive objects found in the field \citep{Zwart2010}.  To begin with, \citet{Blaauw1961} placed a minimum threshold on space velocity at 40 km s$^{-1}$, but over time this has moderated to 30 or 25 km s$^{-1}$ \citep{Hoogerwerf2001,Zwart2000}. The favoured mechanisms for ejection are kicks within binaries arising from supernova explosions or dynamical intra-cluster interactions \citep[see discussion in][]{Hoogerwerf2001}.  Both mechanisms can produce space velocities of up to 200 km s$^{-1}$.  The binary frequency among runaway stars is still uncertain: based on examination of a bright sample dominated by O stars, \citet{Mason2009} concluded that the frequency is lower among confirmed runaways than in clusters, although this is contradicted by \citet{Chini2012}. 

  To better understand the constraints on how and where the most massive stars can form, it is helpful to properly characterize the field population and establish the relative numbers of runaway objects and stars likely to have formed in situ.  We are in a better position to do this if we remove current biases in the on-sky two-dimensional distribution of known O stars -- existing Galactic compilations are dominated by the local volume (to $\sim$ 2 kpc) and, on longer scales, by the contents of recognised open clusters \citep[e.g.][]{Sota2014}.  Now, we can link up what we already know on the few-arcminutes scale around massive Galactic clusters, with the wider field on the scale of degrees, thanks to recent wide-area digital photometric surveys.   %As this step is taken, the current reliance on arbitrary kinematic cuts can be phased out along with distinctions between 'cluster' and 'field' objects.  
%A useful contribution to this will come from sharpening up the presently very fuzzy third dimension.  This will begin soon, as the imminent release of advanced astrometry from ESA's Gaia mission (Gaia DR2) is put to use in delivering constraints on both the parallaxes and proper motions of massive stars.  

We begin to explore this here via a worked example of the massive star content in the wider environment around Westerlund 2, a compact cluster near the tangent of the Carina Arm.  We present follow up spectroscopy and radial velocity measurements of two confirmed very hot O stars, clearly exterior to the main clustering.  For these, we clarify their basic stellar parameters and ask whether they are potentially recent ejections. Westerlund 2 is one of a limited number of dense clusters in the Milky Way estimated to be more massive than $10^4$ M$_{\odot}$ -- recently, \cite{Zeidler2017} have obtained $(3.6\pm0.3)\times10^4$ M$_{\odot}$.  About the oldest age estimated for it is 2.5~Myrs \citep{Rauw2007}: this cluster is young enough that the first supernova explosion is probably yet to happen. If so, the mechanism for creating runaway stars is limited to dynamical interaction.

In two previous papers \citep[][hereafter MS-I and MS-II]{MMS2015, MMS2017}, we presented validated blue selections of OB stars from the VST
%\footnote{The VST is the VLT Survey Telescope, and the VLT is the Very Large Telescope, a set of 4 8-m telescopes operated by the European Southern Observatory} 
Photometric H$\alpha$ Survey of the Southern Galactic Plane and Bulge \citep[VPHAS+,][]{Drew2014} across the Carina region.  Our method of selection incorporated fits to merged VPHAS+ optical and 2MASS NIR photometry that provided a high-quality characterisation of the extinction towards the selected objects. The typical precisions achieved were $\sim$0.1 in each of $A_0$, the monochromatic extinction in magnitudes at 5495~\AA , and $R_V$, the ratio of total to selective extinction.  This extra information provides a start on teasing out the relationship between a young cluster and its wider environment.  The fits also gave a rough constraint on effective temperature that permits the efficient selection of likely O stars.  For Westerlund 2 and its hinterland, we have all these data. 

%The new catalogue of candidate OB stars included both discoveries close to the centre of Westerlund~2 itself and a handful of similarly reddened O stars scattered around it at offsets of between 10 and 40 arcminutes.  The latter may have been ejected from the cluster or they may be evidence of a wider star-forming event within which Westerlund 2 is the most prominent feature.  To settle this, spectroscopic follow up is required.

In a certain respect the methods applied here and the questions addressed parallel the work of \citet{Lamb2016}, with the difference that the 'stellar field' here is a sky region around one massive Galactic cluster, while \citet{Lamb2016} investigated the Small Magellanic Cloud OB population, selecting objects at least 28~pc distant from any other OB candidates.  They found evidence from their sample of 399 stars that around a third are runaways and up to half may have formed in extreme isolation.  Here we consider stars that are at least $\sim$15 pc from Westerlund 2, and consider in detail a projected on-sky region of 130$\times$130 pc$^2$.   The new feature is that we focus on the relation between a specific dense massive cluster, just a few kiloparsecs away, and the scatter of massive stars near it.  We also deploy extinction as a first, crude distance proxy to limit depth in the third dimension. 

This paper is organised as follows.  First, we extract from the larger catalogues presented by MS-I and MS-II a set of high-confidence O stars within a region of 1.5$\times$1.5 sq.deg region centred on Westerlund 2 (hereafter Wd2,  (section~\ref{sec:sample}).  We then introduce additional spectroscopy obtained of two of the hottest O stars in the set (section~\ref{sec:obs}), and describe an analysis that yields improved stellar parameters including an estimate of mass (section~\ref{sec:parms}).  The results of radial velocity measurements for these stars are then presented (section~\ref{sec:rvs}).  The results are considered together in section~\ref{sec:appraisal} with optical and near-infrared survey photometry in an initial appraisal of the links between the dispersed O stars in the region and Wd2.  %This leads us to demote a number of candidates as associated or ejected on a variety of grounds, leaving a group of $2 + 7$ stars of particular interest.  
We then collect and analyse the newly-released Gaia DR2 proper motions for the whole sample (Section~\ref{sec:gaia}): this confirms the two stars with spectroscopy as ejections from Wd2, along with up to 9 more.  The paper ends with a discussion of the results in the context of relevant models for O-star dispersal into the field \citep{Fujii2011, Fujii2012, Lucas2018}.

\section{Westerlund 2 and the O stars around it}
\label{sec:sample}

\begin{figure}
\begin{center}
\includegraphics[width=0.8\columnwidth]{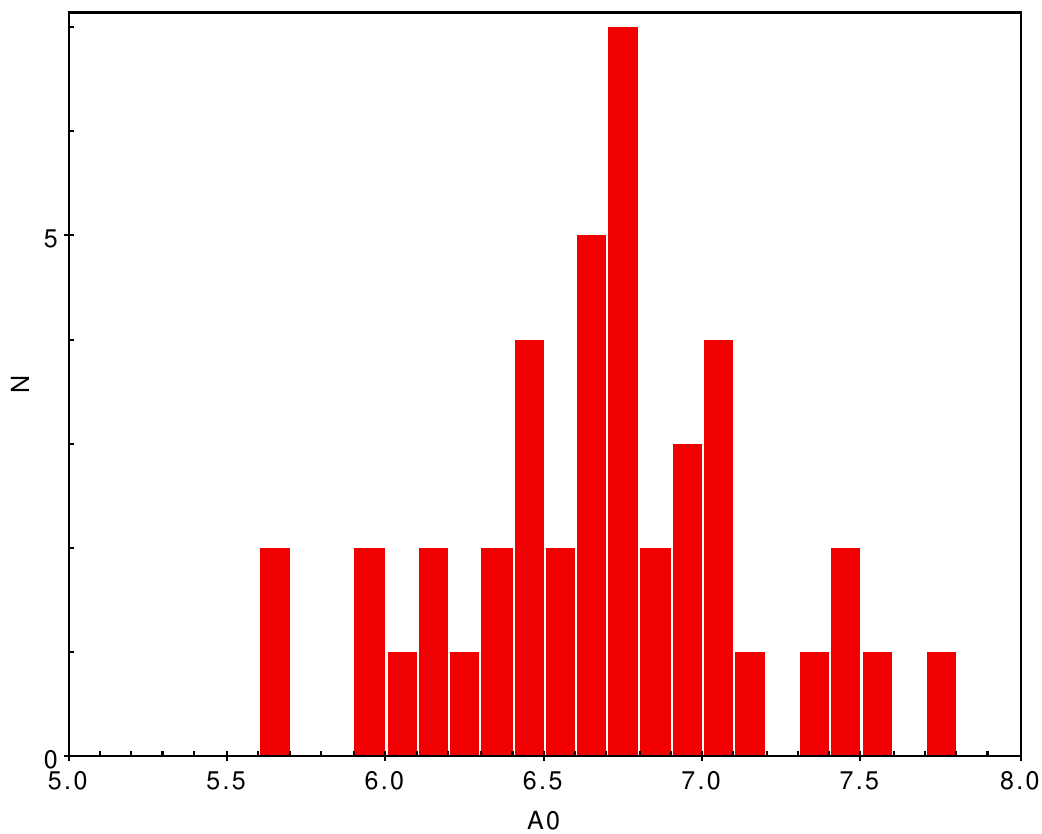}
\includegraphics[width=0.8\columnwidth]{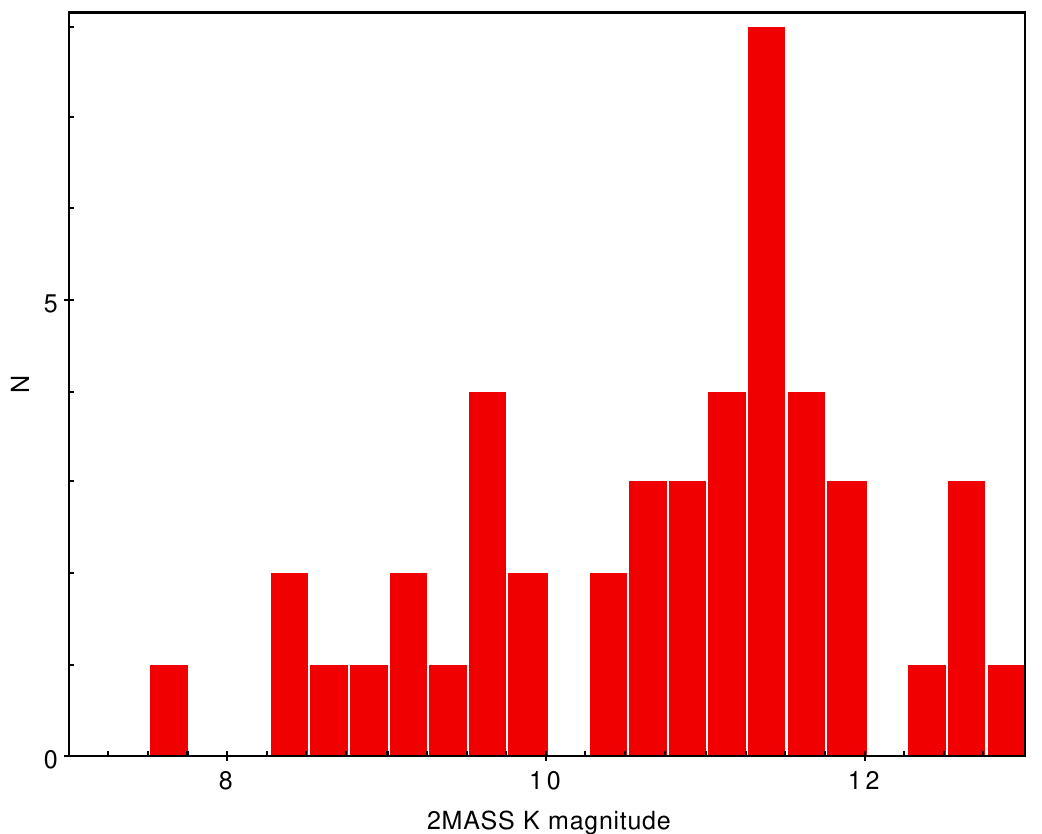}
\caption{
Top: the distribution of extinction, $A_0$ (magnitudes) for 43 of the 46 MS-II catalogue OB stars located on the sky within 6 arcmin of the centre of Wd2. The plotted range is restricted to $5 < A_0 < 8$ to bring out the main grouping.  The remaining 3 stars are at greater extinction ($8 < A_0 < 9$). Bottom: the distribution of $K$ magnitudes, for all 46 stars within 6 arcmin of Wd2 centre.  
}
\label{fig:members}
\end{center}
\end{figure}

\subsection{The centre of Wd2 and the cluster distance}

Expressed in Galactic co-ordinates, the approximate centre of Wd2 was given by \cite{Moffat1975} as $\ell = 284.3^{\circ}$, $b = -0.3^{\circ}$. We shall use $\ell = 284^{\circ}.27$, $b = -0^{\circ}.334$ as our central reference position - it is located in the main cluster about 0.3 arcmin from the very massive WR 20a binary system.  The box we place around this captures the ranges, $283.5^{\circ} < \ell < 285.0^{\circ}$, and $-1.0^{\circ} < b < 0.5^{\circ}$. 

Studies that have focused on the stellar content of Wd2 have typically examined the inner few arcminutes, or few parsecs [\citet{Moffat1991} and \citet{Zeidler2017} represent bookends in time to the several papers that have been published].  The X-ray data collected by \citet{Tsujimoto2007} were exceptional in spanning a region of 17$\times$17 sq.arcmin (backed up by NIR photometry over the inner 8.3$\times$8.3 sq.arcmin). This permitted these authors to argue that the stellar over-density associated with Wd2 may have a radius of 6--7 arcmin (or $\sim$10 pc), and they identified 14 candidate early-type stars more than 3 arcmin away from the cluster centre.  In the more recent catalogue created by MS-II, 46 objects, fitting successfully as OB stars, fall within a circle of radius 6 arcmin around the cluster centre: most are already known in the literature to be OB stars.  The list of 46 is not complete as some objects cannot be extracted in the brilliant cluster core (as imaged in the VPHAS$+$ survey) due to source crowding.  But, for present purpose, it is large enough subset to portray typical Wd2 cluster-member properties reliably.  Specifically, in figure~\ref{fig:members}, the extinction and 2MASS K magnitude distributions near cluster centre are shown.

Estimates of the distance to Wd2, have ranged from 2.8 kpc \citep{Ascenso2007} up to around 8 kpc \citep{Rauw2011}. But there has been some convergence recently on 4--6 kpc among optical stellar photometric studies that have paid attention to the significant variation of the extinction law away from the typical $R_V = 3.1$ version \citep[][and MS-I]{Vargas2013,Hur2015,Zeidler2015}.  Studies of the molecular and ionized interstellar medium have favoured a somewhat longer sightline of 5--7 kpc \citep{Dame2007,Furukawa2009,Benaglia2013}.  

The recent availability of Gaia DR2 does not help us here since the current global astrometric solution has been reported by \cite{Lindegren2018} as carrying a systematic uncertainty of up to 0.1 mas: given that parallaxes for objects in Wd2 are mostly under 0.25 mas, it is clear that the data are not yet good enough to provide reliable and precise distance estimates either to individual objects or Wd2 as a whole.  The alternative of deriving a kinematic distance from the proper motions (see Section~\ref{sec:gaia}) is also hindered by the uncertain peculiar motions of young clusters relative to mean rotation \citep{Reid2014}, and the still uncertain knowledge of the mean disc rotation as a function of Galactocentric radius \citep[see e.g.][]{Huang2016,Harris2018,Kawata2018}. 

In order to place the results of this study onto an absolute physical scale we need a working distance.  For this purpose we adopt 5~kpc, roughly in the mid range of the recent optical photometric work -- and at the bottom end of the ISM estimates.  To keep in mind how distance affects derived quantities, we adopt error bounds of $\pm$1 kpc.  Given the accumulation of work on Wd2, a distance of {\em less than} 4 kpc appears relatively unlikely -- see e.g. \cite{Vargas2013} on the likely problems with the \cite{Ascenso2007} result.  This will be important when, in Section~\ref{sec:gaia}, use is made of Gaia DR2 proper motions: a minimum of 4 kpc puts a useful lower limit on transverse speeds deduced from them. Similarly, with $D \geq 4$~kpc, the 1.5$\times$1.5 sq.deg sky region we consider corresponds to a projected area of at least 100$\times$100 pc$^2$.  The 6 kpc upper bound is of less significance, but if it is exceeded, it puts some strain on the stellar atmospheric analysis of Section~\ref{sec:parms}.

%To help make a choice, we turn to the second release of Gaia data \citep[Gaia DR2,][]{Brown2018}.  It includes astrometry of a precision that is limited for absolute measurements to around $\sim$0.1 mas yr$^{-1}$ by the current state of the global astrometric solution \citep{Lindegren2018}.  Given that parallaxes for objects in Wd2 are mostly under 0.25 mas, it is clear that this is not a route to an improved distance estimate: the systematic error would permit anything from $\sim 3$ to almost 10 kpc.   But the proper motions (PMs) are much more informative since, for most of the objects plotted in Figure~\ref{fig:members}, their magnitudes are $\sim$5--7 mas yr$^{-1}$. Hence it is worth determining the median cluster proper motion, from our selection of highly credible cluster members, and assessing how it translates into a kinematic distance. 
%The insight we gain from consideration of the PM kinematic distance is that its uncertainty approaches a kiloparsec, whilst it seems that a distance of between 6 and 7 kpc is favoured.  We adopt 5 kpc as a {\em working} distance -- in the middle of the estimates based on stellar photometry -- whilst very much aware that it could be too low.  

\subsection{The cluster hinterland}
\label{sec:hinterland}

\begin{table*}
\caption{Properties of candidate OB stars within the region of study, more than 10 arcminutes away from the core of Wd2.  Part (a) of the table contains objects with extinctions overlapping those typical of the cluster, while part (b) lists the higher-extinction candidates ($A_0 > 7.5$).  The estimated extinction parameters, $A_0$ and $R_V$, and effective temperatures, $T_{\rm eff}$, given in columns 5 -- 7, are drawn from the MS-II catalogue.  Effective temperatures estimated from photometry with typical errors in the region of 5000~K, are printed in italic type, while those derived from spectroscopy are in normal type and are accompanied by their formal errors.  No $T_{\rm eff}$ estimate is given for either WR 20aa or WR 20c, as the fitting method is, in these cases, unreliable, and only delivers approximate extinction data.  Column 8, $\theta_r$, specifies the angular separation of each object from the Wd2 reference position at $\ell = 284^{\circ}.27$, $b = -0^{\circ}.334$.  The remark, "$A_0 \sim 8$ BG?", in the comments column identifies candidate members of a more reddened, background association (see section~\ref{sec:unrelated}).  Columns 2 and 3 give positions in Galactic coordinates -- alternative RA and Dec (J2000) positions are given in the Appendix, along with the names of cross-matched Gaia DR2 sources. 
} 
\begin{center}
\begin{tabular}{cccrccccl}
\hline
MS-II \# & \multicolumn{2}{c}{Galactic Coordinates} & 2MASS K & A$_0$ & R$_V$ & T$_{eff}$ & $\theta_r$ & comment \\
  & $\ell^{\circ}$ & $b^{\circ}$ & (mag) & (mag)  &    & (kK) & (arcmin) & \\
\hline
\multicolumn{5}{l}{(a) $5.5 < A_0 < 7.5$} & & & & \\ 
\hline	   
0673 &  283.505754 & $-$0.538484 & 11.75$\pm$0.02 & $5.73_{-0.10}^{+0.09}$ & $3.72_{-0.07}^{+0.07}$ & {\it 29} & 47.47 &  \\
0693 &  283.536501 & $-$0.426998 & 10.57$\pm$0.02 & $6.73_{-0.10}^{+0.09}$ & $4.34_{-0.08}^{+0.09}$ & {\it 30} & 44.36 & emission line star \\
0708 &  283.561722 & $-$0.979383 &  7.94$\pm$0.03 & $6.84_{-0.09}^{+0.08}$ & $3.77_{-0.06}^{+0.06}$ & {\it 31} & 57.49 & NGC 3199/foreground \\
0893 &  283.842822 & $-$0.705926 & 10.87$\pm$0.03 & $6.81_{-0.06}^{+0.05}$ & $3.80_{-0.05}^{+0.05}$ & {\it 41} & 33.98 & \\
0986 &  283.937827 & $-$0.830267 & 11.56$\pm$0.05 & $7.40_{-0.10}^{+0.10}$ & $4.07_{-0.07}^{+0.07}$ & {\it 28} & 35.83 & $A_0 \sim 8$ BG? \\
0987 &  283.939130 & $-$0.910605 & 11.93$\pm$0.02 & $7.33_{-0.10}^{+0.09}$ & $3.99_{-0.06}^{+0.06}$ & {\it 29} & 39.89 & $A_0 \sim 8$ BG? \\
0994 &  283.947686 & $-$0.489082 & 10.29$\pm$0.02 & $5.58_{-0.09}^{+0.08}$ & $3.79_{-0.07}^{+0.08}$ & {\it 31} & 21.46 &  \\
1046 &  284.046887 & $+$0.425467 &  9.14$\pm$0.02 & $6.20_{-0.08}^{+0.06}$ & $3.57_{-0.05}^{+0.06}$ & {\it 35} & 47.49 & \\
1102 &  284.142129 & $-$0.188894 &  9.62$\pm$0.02 & $6.28_{-0.07}^{+0.05}$ & $3.64_{-0.05}^{+0.05}$ & {\it 38} & 11.60 & \\
1133 &  284.193343 & $-$0.132585 &  9.93$\pm$0.02 & $7.01_{-0.07}^{+0.05}$ & $3.72_{-0.05}^{+0.05}$ & {\it 38} & 12.93 & \\
1164 &  284.227820 & $-$0.687882 &  9.65$\pm$0.02 & $6.40_{-0.08}^{+0.06}$ & $3.91_{-0.06}^{+0.07}$ & $34.3_{-0.6}^{+0.5}$ & 21.38 & \\
1236 &  284.276442 & $-$0.163506 & 10.04$\pm$0.02 & $7.29_{-0.10}^{+0.09}$ & $3.53_{-0.05}^{+0.05}$ & $41.8_{-3.2}^{+6.2}$ & 10.24 & \\
1273 &  284.298853 & $-$0.519768 &  9.52$\pm$0.02 & $6.67_{-0.09}^{+0.06}$ & $3.87_{-0.06}^{+0.06}$ & $40.6_{-0.8}^{+2.5}$ & 11.28 & X-shooter target  \\
1308 &  284.331634 & $-$0.583555 &  8.39$\pm$0.03 & {\it 5.6} & {\it 4.3} & -- & 15.42 & WR 20aa  \\
1338 &  284.378387 & $+$0.009138 &  9.41$\pm$0.02 & $6.17_{-0.07}^{+0.05}$ & $3.58_{-0.05}^{+0.05}$ & $42.5_{-0.8}^{+0.7}$ & 21.59 & X-shooter target \\
1356 &  284.402022 & $-$0.548087 & 10.34$\pm$0.02 & $5.68_{-0.06}^{+0.05}$ & $4.05_{-0.07}^{+0.07}$ & $39.0_{-0.4}^{+0.3}$ & 15.09 &  \\
1374 &  284.418116 & $-$0.927350 & 10.54$\pm$0.02 & $5.61_{-0.09}^{+0.08}$ & $3.75_{-0.07}^{+0.07}$ & $38.6_{0.4}^{+0.5}$ & 36.69 & \\
1550 &  284.638801 & $-$0.499403 & 10.43$\pm$0.04 & $6.87_{-0.10}^{+0.10}$ & $3.98_{-0.07}^{+0.07}$ & $36.4_{1.1}^{+1.5}$ & 24.25 & \\
1567 &  284.657774 & $-$0.775554 &  9.89$\pm$0.02 & $5.55_{-0.09}^{+0.08}$ & $3.71_{-0.07}^{+0.07}$ & $34.3_{-0.3}^{+0.2}$ & 35.26 & in own cluster \\
\hline
\multicolumn{5}{l}{(b) $A_0 > 7.5$} & & & &  \\
\hline
0685 & 283.527633 & $-$0.860257 & 11.54$\pm$0.03 & $8.33_{-0.14}^{+0.08}$ & $3.70_{-0.05}^{+0.05}$ & {\it 32} & 54.60 &  \\
0785 & 283.684167 & $+$0.417875 & 12.36$\pm$0.03 & $7.81_{-0.15}^{+0.09}$ & $3.64_{-0.05}^{+0.06}$ & {\it 31} & 57.19 &  \\
0826 & 283.744780 & $-$0.630411 & 10.34$\pm$0.02 & $9.84_{-0.13}^{+0.06}$ & $4.05_{-0.05}^{+0.05}$ & {\it 35} & 36.19 & \\
0879 & 283.828965 & $-$0.736854 & 10.71$\pm$0.02 & $7.77_{-0.08}^{+0.06}$ & $3.89_{-0.05}^{+0.05}$ & {\it 36} & 35.84 & $A_0 \sim 8$ BG?  \\
0881 & 283.829449 & $-$0.595493 & 10.83$\pm$0.03 & $8.28_{-0.12}^{+0.11}$ & $3.91_{-0.05}^{+0.05}$ & {\it 28} & 30.74 & $A_0 \sim 8$ BG? \\
0896 & 283.848489 & $-$0.848510 &  9.60$\pm$0.02 & $8.03_{-0.08}^{+0.06}$ & $4.07_{-0.05}^{+0.06}$ & {\it 35} & 39.91 & $A_0 \sim 8$ BG? \\
0904 & 283.858744 & $-$0.932254 & 12.39$\pm$0.03 & $7.72_{-0.07}^{+0.05}$ & $3.88_{-0.05}^{+0.05}$ & {\it 39} & 43.56 & $A_0 \sim 8$ BG? \\
0918 & 283.873029 & $-$0.917056 & 10.86$\pm$0.02 & $8.15_{-0.05}^{+0.05}$ & $3.93_{-0.05}^{+0.05}$ & {\it 40} & 42.32 & $A_0 \sim 8$ BG? \\
0919 & 283.873213 & $-$0.910814 & 11.25$\pm$0.03 & $7.76_{-0.10}^{+0.09}$ & $3.88_{-0.05}^{+0.06}$ & {\it 30} & 42.01 & $A_0 \sim 8$ BG? \\
0921 & 283.875873 & $-$0.910293 & 10.14$\pm$0.03 & $8.35_{-0.10}^{+0.08}$ & $3.95_{-0.06}^{+0.06}$ & {\it 31} & 41.89 & $A_0 \sim 8$ BG? \\
0925 & 283.879042 & $-$0.917201 & 10.44$\pm$0.04 & $8.37_{-0.06}^{+0.06}$ & $3.95_{-0.05}^{+0.05}$ & {\it 41} & 42.13 & $A_0 \sim 8$ BG? \\
0930 & 283.885189 & $-$0.907088 & 11.77$\pm$0.05 & $8.33_{-0.08}^{+0.06}$ & $3.91_{-0.06}^{+0.06}$ & {\it 39} & 41.42 & $A_0 \sim 8$ BG? \\
0931 & 283.885209 & $-$0.912074 &  9.96$\pm$0.03 & $9.34_{-0.09}^{+0.06}$ & $3.75_{-0.04}^{+0.04}$ & {\it 37} & 41.67 & \\
0934 & 283.885885 & $-$0.960334 & 12.69$\pm$0.04 & $7.56_{-0.13}^{+0.11}$ & $4.08_{-0.07}^{+0.07}$ & {\it 29} & 44.08 & $A_0 \sim 8$ BG? \\
0938 & 283.889741 & $-$0.563901 & 11.95$\pm$0.03 & $8.59_{-0.19}^{+0.12}$ & $4.09_{-0.06}^{+0.06}$ & {\it 30} & 26.66 & $A_0 \sim 8$ BG? \\
0953 & 283.903648 & $+$0.372327 & 11.73$\pm$0.04 & $8.26_{-0.19}^{+0.13}$ & $3.70_{-0.06}^{+0.06}$ & {\it 29} & 47.74 & \\
0958 & 283.908449 & $-$0.903378 & 11.63$\pm$0.02 & $7.60_{-0.10}^{+0.08}$ & $3.88_{-0.06}^{+0.06}$ & {\it 31} & 40.47 & $A_0 \sim 8$ BG? \\
0965 & 283.917002 & $-$0.804651 & 12.43$\pm$0.04 & $7.79_{-0.10}^{+0.08}$ & $4.04_{-0.07}^{+0.07}$ & {\it 34} & 35.30 & $A_0 \sim 8$ BG? \\
0968 & 283.917749 & $-$0.964225 & 11.59$\pm$0.03 & $8.44_{-0.14}^{+0.08}$ & $3.86_{-0.05}^{+0.05}$ & {\it 33} & 43.32 & $A_0 \sim 8$ BG? \\
1086 & 284.119503 & $-$0.071365 &  9.65$\pm$0.02 & $9.67_{-0.14}^{+0.07}$ & $3.54_{-0.04}^{+0.04}$ & {\it 35} & 18.16 & \\
1119 & 284.175755 & $+$0.077817 &  9.04$\pm$0.02 & {\it 10.3} & {\it 3.7} & -- & 25.35 & WR 20c  \\
\hline
\end{tabular}
\end{center}
\label{tab:list}
\end{table*}

\begin{figure*}
\begin{center}
\includegraphics[width=1.3\columnwidth]{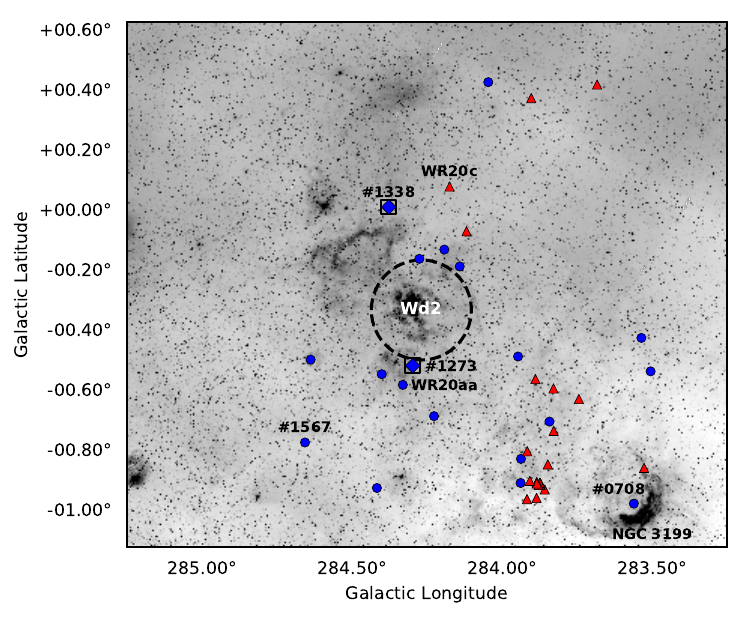}
\caption{
The region around Westerlund 2.  The blue-selected O stars from the catalogue of MS-II, listed in Table~\ref{tab:list}, are superposed on an image of the field constructed from VPHAS+ data \citep{Drew2014}. All the named objects, including the nebula NGC 3199, are discussed later in the text.  For emphasis, the two X-shooter targets, stars \#1273 and \#1338, are diamonds enclosed in boxes.  Stars appearing in the lower-extinction/upper half of table~\ref{tab:list} are drawn as circles and coloured blue, while the red triangles pick out the higher-extinction objects in the lower half of the same table.  The black dashed circle is centred on the reference position, $\ell = 284^{\circ}.27$, $b = -0^{\circ}.334$, and has a radius of 10 arcmin.  The radial velocity comparison stars, MSP 182, MSP 183 and MSP 199 (not marked) are all located in the cluster centre, within an arcminute of the reference position. 
}
\label{fig:map}
\end{center}
\end{figure*}

As a first step to identifying candidate O stars in the immediate field around Wd2, we apply the following criteria.  First, we restrict attention to objects at an angular separation from the cluster reference position that exceeds 10 arcmin: this minimum angular scale maps onto a plane-of-sky distance at 5 kpc of 15 pc -- or the distance travelled in 1~Myr at a projected speed of 15 km s$^{-1}$.  The majority of cluster escapes, with full three-dimensional space velocities of $\sim$30 km s$^{-1}$ or more should lie outside this circle. 

Next, we take from the catalogue of MS-II as our primary sample those stars hot enough to classify as O stars ($\log(T_{eff}) > 4.45$) for which the estimated extinction lies within the range, $5.5 < A_0 < 7.5$ (motivated by the top panel of figure~\ref{fig:members}).  This is aimed at cutting out objects likely to be well into the foreground with respect to Wd2.  We also limit ourselves to objects whose OIR photometry fits with high-confidence to a reddened OB spectral energy distribution ($\chi^2 < 8$, see MS-I).  The one exception we make to this last rule is for the Wolf-Rayet star, WR 20aa, for which $\chi^2 \simeq 13$.  Finally, we drop from consideration any relatively faint objects $K > 12.5$ (cf. the bottom panel of figure~\ref{fig:members}).

The 19 objects satisfying all the above criteria are set out in the top half of Table~\ref{tab:list} along with their 2MASS $K$ magnitudes, and the extinction parameters and effective temperatures reported by MS-II.  In most cases, the tabulated effective temperature estimate is based on the fit to VPHAS$+$ and 2MASS photometry.  However, for some, we have more reliable spectroscopically-based estimates (see MS-II for further discussion).  

We retain, as a distinct group, candidate objects with extinctions exceeding the $A_0 = 7.5$ limit imposed on the main sample: we view them as potentially in the cluster's background, and thus less likely to be associated with it. This applies the broad principle that extinction rises with distance (see the extensive discussion of this and how the region around Wd2 is challenging in MS-II). The same minimum effective temperature, the same angular separation from Wd2, and the same confidence cut are imposed, giving a group of objects that is in the mean $\Delta K = 0.7$ fainter (and no object is fainter than $K = 12.7$).  The data on this set of 21 stars are presented in the lower half of Table~\ref{tab:list}.  The highest-extinction object in this group, is WR20c ($A_0 \simeq 10.3$): this object has been claimed as one of a runaway pair from Wd2 by \citet{RomanLopes2011}.  WR 20aa, in the upper half of the table, is the other star making up the proposed pair. 

How both groups of objects are distributed on the sky is shown in figure~\ref{fig:map}.  Our initial hypothesis is that the first candidate group are the best options for physical proximity to Wd2: whatever the distance to Wd2 is, these stars should be at essentially the same distance until the balance of evidence gives reasonable doubt. This is not anticipated for the second group.  We return to this in the closing discussion (Section~\ref{sec:discussion}).

%Table~\ref{tab:list} includes the positions of all object in Galactic co-ordinates as given by MS-II.  In the Appendix, a further table is provided that gives full MS-II object names, positions in RA,Dec, and the names of the Gaia DR2 cross-match objects. 

\section{Observations}
\label{sec:obs}

The multi-colour photometry and low resolution spectroscopy underpinning this study was presented and described by MS-II.  Here we complete the picture with a description of higher-quality observations obtained since, of two of the hottest O-stars in the circum-cluster environment of Wd2, using the X-shooter instrument on ESO's Very Large Telescope (VLT) in April/May 2015. 

\subsection{Choice of targets for higher resolution spectroscopy}

\begin{table*}
\caption{Positions and properties of the X-shooter targets.}
{\centering
\begin{tabular}{lccccl}
\hline  
Target name & RA (J2000) & Dec (J2000) & \multicolumn{2}{c}{$g$} & Spectral type and heliocentric \\
   & & &        &           & RV from Rauw et al (2011) \\
   & & & Vega mag. & source    & \\
\hline
\#1273 & 10 23 26.43 & $-$57 56 03.68 & 15.56 & MS-II &  \\
\#1338 & 10 26 03.10 & $-$57 31 43.06 & 14.99 & MS-II &  \\
MSP 182 & 10 23 56.18 & $-$57 45 30.00 & 15.33 & MS-II & O4V-III((f)), RV $= 30.5$ km s$^{-1}$\\
MSP 199 & 10 24 02.65 & $-$57 45 34.33 & 14.79 & VPHAS$+$ DR2 & O3-4V, RV $= 30.6$ km s$^{-1}$ \\
MSP 183 & 10 24 02.36 & $-$57 45 30.59 & 15.41 & VPHAS$+$ DR2 & O3V((f)), RV $= 23.9$ km s$^{-1}$ \\
\hline
\end{tabular}
}
\label{table:targets}
\end{table*}

\begin{table*}
\caption{Journal of X-shooter observations.  The given observation timings correspond to mid-observation times.}
\begin{center}
\begin{tabular}{ccccc}
\hline
Target name & \multicolumn{2}{c}{Galactic Coordinates} & Date (MJD) & UVB exposure time (secs) \\
   & $\ell$ (deg.) & $b$ (deg.) &  & \\
\hline
\#1338 & 284.29885 & $-$0.51977 & 57132.002615 & 2 x 600 \\
       &           &          & 57135.108884 & 2 x 600 \\
       &           &          & 57142.049693 & 2 x 600 \\ 
\#1273 & 284.37839 & $+$0.00914 & 57142.994510 & 2 x 900 \\
       &           &          & 57144.992392 & 2 x 900 \\
       &           &          & 57155.986232 & 2 x 900 \\
       &           &          & 57158.986483 & 1 x 900 \\
MSP 182 & 284.26045 & $-$0.33575 & 57132.032011 & 2 x 800 \\
MSP 199 & 284.27326 & $-$0.32911 & 57143.095241 & 2 x 800 \\
MSP 183 & 284.27221 & $-$0.32859 & 57156.013090 & 2 x 600 \\
\hline
\end{tabular}
\end{center}
\label{table:observations}
\end{table*}

MS-I identified a number of potential O-star ejections from Wd2, based on similarity of extinction to those measured in the cluster itself: 14 objects were found at separations up to 80 arcmin within the extinction range $5.8 < A_0 < 7.2$ (i.e. to within $1\sigma$ of the measured mean for Wd2, see Table 8 in MS-I).  Some of these were included as targets in the low-resolution spectroscopic follow-up programme reported by MS-II.  Two objects, named \#916 and \#646 by MS-I, were confirmed as very hot O stars: fits to the low resolution spectra obtained indicated effective temperatures of $\sim$42.5 kK and $\sim$40.6 kK respectively, and surface gravities compatible with the main sequence.  Here, they are named according to the superseding MS-II catalogue as VPHAS-OB1-01338 and VPHAS1-OB1-01273, or \#1338 and \#1273, for short.  

Since these first spectra were not well enough resolved for radial velocity measurement, we obtained X-shooter observations over a spread of dates in April and May 2015 that were.  To place the new spectra in the Wd2 context, we also obtained observations of three O stars located in the heart of the cluster, that have comparable effective temperatures and known radial velocities.  They were selected from the study by \citet{Rauw2011} as suitable reference objects since they presented no obvious radial velocity changes due to binary motions.  They are MSP 182, 183 and 199 with mean measured heliocentric radial velocities of 30.5, 23.9 and 30.6 km s$^{-1}$ (as obtained from Gaussian fits to the HeII 541.2 nm line).  Some of the properties of these stars are set down in Table~\ref{table:targets}.

\subsection{X-shooter observations}

The X-shooter spectrograph delivers a spectral resolution of $R = 11000$ in the blue/optical wavelength range where the strongest photospheric absorption lines seen in O star spectra are located.  The target stars \#1338 and \#1273 were observed in queue mode 3 and 4 times each, over time spans of 10 and 17 days respectively, with the reference objects interspersed.  A journal of all the blue/optical observations obtained per target is given as Table~\ref{table:observations}.  The seeing at the time of observation was generally not more than 1 arcsec, but always exceeded the slit width of 0.5 arcsec.   The data were extracted into one-dimensional spectra using the {\sc REFLEX} package \citep{Freudling2013}, using default settings of the routines appropriate to the slit/stare mode.  In the case of MSP~183, the traces of two fainter stars are visible in the image frames towards one end of the 11 arcsec slit: since the sky subtraction uses median values sampled from $2\times4$~arcsec along the slit, these will have had minimal impact on the final spectrum.

%{\bf Thomas to write - we will need to know how the spectrum was extracted from the 11 arcsec slit: in MSP 199 and 183 there is a concern that there may be some contaminating flux from companion objects.  Can we quantify?}

\section{Results}

%In our further discussion we will adopt as radial velocities (relative to MSP 182) the rounded off values of $-33$ and $-16$ km s$^{-1}$.

\subsection{The stellar parameters of \#1338, \#1273 and the three comparison objects}
\label{sec:parms}

Here, we present the results of fitting NLTE model-atmosphere line profiles to selected transitions in the blue X-shooter spectra.  Fits have been applied to the RV control stars (MSP 182, 199 and 183) as well as to \#1338 and \#1273 so that we can place our results in the context of previous work.

\begin{figure*}
\begin{center}
\includegraphics[width=1.7\columnwidth]{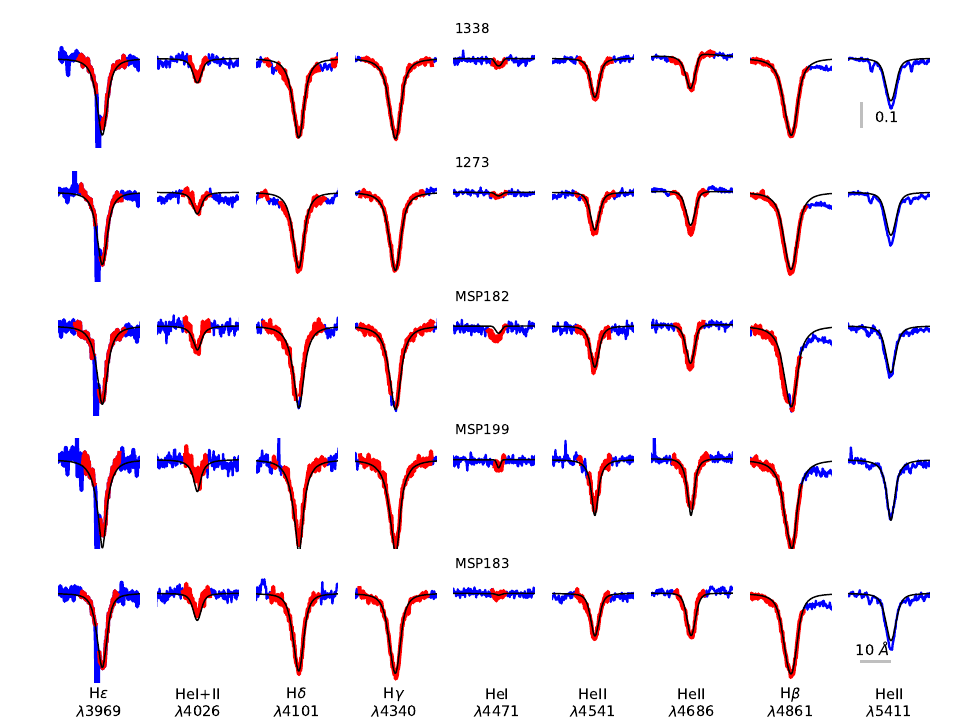}
\caption{
Final fits to the observed target and control stars. We represent in red the part of the observed spectra that has been used for the analysis, in blue parts of the observed spectra not used for the analysis and in black the theoretical spectra, convolved with the rotational and macroturbulent velocities and the instrumental profile (assumed gaussian). The gray horizontal and vertical lines give the axes scales
}
\label{fig:fits}
\end{center}
\end{figure*}

The fitting methodology is as described in \citet{sabin14} and \citet{holgado18}. Briefly, we first determine the rotational and so-called {\it macroturbulent} velocities with the {\sc IACOB-BROAD} tool \citep[see][]{sh14}, which uses both the Fourier transform and goodness-of-fit methods. With these values, we then use the grid-based automatic tool {\sc IACOB-GBAT} \citep{ssimon11} to determine the stellar parameters. {\sc IACOB-GBAT} finds them using an extensive grid of model atmospheres calculated with the code {\sc FASTWIND} \citep{santolaya97,puls05} covering a large range of stellar model parameters. Six model parameters are varied in the grid: the effective temperature $T_{\rm eff}$; stellar gravity $\log g$; helium abundance by number relative to hydrogen $Y_{\rm He}$; the microturbulence; the exponent of the wind velocity field -- assumed to be a $\beta$ law; the wind-strength parameter $Q= {{\dot{M}}/(R v_\infty)^{1.5}}$, where $\dot{M}$ is the mass-loss rate, $R$ the stellar radius and $v_\infty$ the wind terminal velocity. A $\chi^2$ analysis is carried out over a grid of nearly 200\,000 models (see the above references for more details). Final fits can be seen in Fig.~\ref{fig:fits}.

The results for effective temperature, surface gravity, helium abundance and wind strength parameter are set down in table~\ref{tab:parameters} together with their uncertainties\footnote{the uncertainty in the surface gravity of star $\#1338$ has been increased to $\pm$0.10 instead of the formal one of $\pm0.05$, following the discussions in \citet{sabin17} and \citet{holgado18}}.  The errors on microturbulence and the wind $\beta$ exponent are degenerate, i.e., there is not enough information in the observed spectrum to determine them, and so we do not  specify them.  At a first glance, all the derived surface gravities are compatible with the dwarf main sequence status: \citet{MSH2005} give $\log g = 3.92$ (as compared with $<3.8$ for the hottest giants and supergiants), and the recent work by \citet{holgado18} also indicates that these gravities correspond to luminosity class V stars.  Compared to the fits to the lower spectral resolution AAOmega results reported by MS-II, the effective temperatures of \#1338 and \#1273 are raised (to $\sim 45$~kK) but not by amounts incompatible with the uncertainties.

Because there is only one epoch of observation available, the determinations for the control stars are less precise, reflecting the lower total observed counts.  However, it is striking that all the derived parameters for MSP 182 \citep[O4V,][]{Rauw2011} nearly match those for \#1338 and \#1273 (although the fit to He\,I 447.1 nm is poorer).  The absolute visual magnitudes of the three objects are consistent with the value of -5.50 assigned to O4V by \citet{MSH2005}, given the likely uncertainty in this quantity of $\sim$0.15.  The effective temperatures are all a little high, by $\sim 1.5-2.0$~kK, relative to the \citet{MSH2005} and \citet{holgado18} scales. However, the number of stars with spectral type O4 and earlier considered by these authors is very small and the dispersions relatively large. The derived helium abundances, $Y_{\rm He}$, are expressed by number relative to hydrogen and they too agree and are normal.  The wind strength parameter log$Q$ is in the upper range of the values given by \citet{holgado18} for dwarfs, as corresponds to their early spectral types. Taken as a whole, these results suggest that all three objects are plausibly normal abundance O4V stars at much the same distance from us: in terms of relative error the $\sim$0.3 mag dispersion in derived $M_V$ is inside the standard deviation of 0.4 mag specified for dwarfs by \cite{MSH2005}.
%Usefully, MSP 182 is situated in a lower density part of the central cluster and is the least troubled by potential on-slit contamination from neighbours.       

The other control stars, MSP 199 and MSP 183, lie in the more crowded heart of Wd2 and are more at risk of contamination, both photometrically and spectroscopically.  In the tabulated results of the Hubble Space Telescope (HST) analysis by \citet{Vargas2013}, respectively two and six close companions to MSP 199 and MSP 183 are listed.  The next brightest companion to MSP 199 is $\sim$2.4 magnitudes fainter in $V$ and at an angular separation of 1.3 arcsec, while for MSP 183 these numbers become $\sim$2.0 magnitudes and 1.5 arcsec.  Inspection of the X-shooter 2-D images indicates negligible contamination (at worst, sky subtraction of MSP 183 is marginally affected).  But there are raised uncertainties in the photometric magnitudes extracted from ground-based overlapping stellar images.  

In recent dedicated photometric studies of Wd2, the cited $V$ magnitudes of MSP 199 and MSP 183 each span a range of $\sim$0.3.  Unfortunately it is not quite as straightforward as adopting what ought to be cleaner HST $V$ magnitudes because of the evidence that the absolute scale applied by \citet{Vargas2013} is slightly faint (see discussion of this point in MS-I).  For both stars, and also MSP 182, we use the \citet{Rauw2007} visual magnitudes since they sit in the midst of the published alternatives.  For the visual extinction of MSP 183, we provisionally adopt and adapt the \citet{Rauw2007} result -- replacing their assumed $R_V = 3.1$ law, by the more suitable $R_V = 3.8$ law.  This gives $A_V \simeq 6.8$.  We caution this is much more uncertain than the extinctions for the other stars since there is no direct measurement of $R_V$.  

\begin{table*}
\caption{Apparent magnitudes and the absolute visual magnitudes derived from them are presented here, along with the fundamental astrophysical stellar parameters derived from fitting NLTE model-atmosphere line profiles to the observations.  The sources consulted for the apparent magnitudes and visual extinctions are given, except for the more crowded and challenging case of MSP 183 (see discussion in text).  The data for MSP 183 affected by the greater uncertainty in $A_V$ are shown in italic font.  The absolute visual magnitudes have been computed for a distance of 5 kpc -- note that these would be 0.5 magnitudes fainter if a distance of 4 kpc were adopted instead.}
\begin{center}
\begin{tabular}{lccccccccc}
\hline  
Star & $V$ & $V$ Ref & $A_V$ & $A_V$ Ref & $M_V$ & $T_{\rm eff}$ & $\log g$ & $Y_{\rm He}$ & log$Q$ \\
   & (mag.) &        & (mag.) &          & (mag.) & (kK) & & & \\
\hline
\#1338 & 13.95 & MS-II & 6.17 & MS-II & -5.71 & 45.2$\pm$0.9 & 3.90$\pm$0.10 & 0.10$\pm$0.01 & $-$12.38$\pm$0.10 \\
\#1273 & 14.49 & MS-II & 6.67 & MS-II & -5.67 & 46.3$\pm$1.6 & 3.92$\pm$0.12 & 0.11$\pm$0.03 & $-$12.70$\pm$0.26 \\
MSP 182 & 14.45 & Rauw et al 2007 & 6.37 & Vargas et al 2013 & -5.41 & 45.5$\pm$3.5 & 3.98$\pm$0.23 & 0.12$\pm$0.05 & $-$12.57$\pm$0.28 \\
MSP 199 & 14.36 & Rauw et al 2007 & 6.40 & Vargas et al 2013 & -5.53 & 49.1$\pm$3.4 & 4.05$\pm$0.15 & 0.15$\pm$0.07 & $-$12.78$\pm$0.34 \\
MSP 183 & 13.57 & Rauw et al 2007 & {\it 6.8} & see text & {\it -6.7} & 49.0$\pm$3.0 & 3.88$\pm$0.12 & 0.16$\pm$0.05  & $-$12.79$\pm$0.33 \\
\hline
\end{tabular}
\end{center}
\label{tab:parameters}
\end{table*}

% ORIGINAL TABLE 6
%\begin{table}
%\caption{Derived stellar radii and masses.  As in table~\ref{tab:parameters}, the less certain data for MSP 183 are shown in italic font.}
%\begin{center}
%\begin{tabular}{lcccc}
%\hline  
%Star & $M_V$ & $F_m$ & $R/R_{\odot}$ & $M/M_{\odot}$ \\
%     & (mag.) &      &             &             \\
%\hline
% Janet's orginal numbers
%\#1338 & $-$5.71 & $-$29.653 & 13.3 & 47 \\
%\#1273 & $-$5.67 & $-$29.685 & 13.1 & 48 \\
%MSP 182 & $-$5.41 & $-$29.658 & 12.1 & 46 \\
%MSP 199 & $-$5.53 & $-$29.767 & 10.5 & 41 \\
%MSP 183 & {\it $-$6.7} & $-$29.758 & {\it 20} & {\it 100}  \\

%Artemio numbers
%\#1338 & $-$5.71 & $-$29.653 & {\bf 13.4} & {\bf 52} \\
%\#1273 & $-$5.67 & $-$29.685 & {\bf 12.9} & {\bf 51} \\
%MSP 182 & $-$5.41 & $-$29.658 & {\bf 11.6} & {\bf 47} \\
%MSP 199 & $-$5.53 & $-$29.767 & {\bf 11.7} & {\bf 56} \\
%MSP 183 & {\it $-$6.7} & $-$29.758 & {\it 20} & {\it {\bf 110}}  \\
%\hline
%\end{tabular}
%\end{center}
%\label{tab:masses}
%\end{table}

%NEW TABLE 6
\begin{table*}
\caption{Derived stellar radii, masses, luminosities and wind momenta at the working distance of 5~kpc. We have adopted an uncertainty of $\pm$0.3 mag for $M_{\rm V}$. As in table~\ref{tab:parameters}, the less certain data for MSP 183 are shown in italic font.  A distance error of $\pm$1~kpc translates into uncertainties of $\pm$0.4 mag in $M_V$, and 20\% in stellar radius.  
}
\begin{center}
\begin{tabular}{lccccccc}
\hline  
Star & $M_V$ & $F_m$ & $R/R_{\odot}$ & $M/M_{\odot}$ & log$L/L_\odot$ & log$\dot{M}$ & log$D_{\rm mom}$ \\
     & (mag.) &      &             &            &        & \\
\hline
\#1338 & $-$5.71 & $-$29.653 & 13.4$^{+2.0}_{-1.7}$ & 53$^{+23}_{-16}$ & 5.83$\pm$0.13 & $-$5.56$\pm$0.14 & 29.22$\pm$0.16 \\
\#1273 & $-$5.67 & $-$29.685 & 12.9$^{+1.9}_{-1.7}$ & 51$^{+24}_{-17}$ & 5.84$\pm$0.14 & $-$5.91$\pm$0.28 & 28.87$\pm$0.29 \\
MSP 182 & $-$5.41 & $-$29.658 & 11.6$^{+1.8}_{-1.6}$ & 48$^{+38}_{-21}$ & 5.72$\pm$0.19 & $-$5.80$\pm$0.30 & 28.97$\pm$0.32 \\
MSP 199 & $-$5.53 & $-$29.767 & 11.7$^{+1.8}_{-1.6}$ & 56$^{+32}_{-20}$ & 5.86$\pm$0.18 & $-$5.97$\pm$0.36 & 28.83$\pm$0.37 \\
MSP 183 & {\it $-$6.7} & $-$29.758 & {\it 20} & {\it 115} & {\it 5.86} & {\it $-$5.66} & {\it 29.25}  \\
\hline
\end{tabular}
\end{center}
\label{tab:masses}
\end{table*}

The collected results for MSP 199 and MSP 183 in table~\ref{tab:parameters} indicate these stars are, at $\sim$49~kK, even hotter than MSP 182.  The sense of this difference is certainly consistent with the earlier O3 spectral types attributed to them by \citet{Rauw2011}.  For types earlier than O4, the extremely weak or absent He\,I lines drive the effective temperatures towards higher values, producing a change in the slope that is still controversial \citep[see][and references therein]{sabin17}. In this respect, the outcome of the model fits is satisfactory.  But the estimate of $M_V$ for MSP 199 is fainter than the -5.85 expected on the main sequence for O3, while for MSP 183 the estimate is appreciably brighter and even too bright for supergiant status \citep[whilst more than 5~kK hotter than viewed as representative for O3I, see][]{MSH2005,holgado18}.  However, their wind strength parameters look normal for early dwarfs. The results for these two objects are thus less convincing than for MSP 182, and oppose each other in terms of the distance that would best suit them -- indeed, our 5 kpc working distance emerges as a rough compromise.  The helium abundances for both MSP 199 and 183 are somewhat elevated, which may imply both are moving away from the zero age main sequence.  This would not be surprising.

\begin{table*}
\caption{Radial velocity determinations.  All are expressed in a frame that places MSP 182 at 0 km s$^{-1}$.  The errors, $\epsilon$, on the individual measurements are the Gaussian fit uncertainties, while the epoch means have been computed weighting each measure by $1/\epsilon^2$.  The errors on the means have been computed from the observed scatter of the data contributing to each mean.}

\label{tab:rvmeasures}
\begin{center}
\begin{tabular}{rccccc}
  & \multicolumn{3}{c}{Cross-correlation RV shifts (km s$^{-1}$)} & Epoch mean & Overall mean \\
  &  wrt MSP 182 & via MSP 199 & via MSP 183 & & \\          
\hline \noalign{\smallskip}
\#1338: epoch 1 & $-$32.2$\pm$2.0 & $-$30.1$\pm$2.5 & $-$28.9$\pm$2.0 & $-$30.4$\pm$2.0 & \\
             2 & $-$30.2$\pm$2.0 & $-$26.2$\pm$2.0 & $-$26.5$\pm$2.0 & $-$27.6$\pm$2.4 & \\
             3 & $-$31.5$\pm$2.0 & $-$28.6$\pm$2.2 & $-$28.3$\pm$1.9 & $-$29.5$\pm$1.4 & $-$29.4$\pm$1.7 \\    
\hline \noalign{\smallskip}
\#1273: epoch 1 & $-$18.2$\pm$2.1 & $-$15.9$\pm$2.0 & $-$15.3$\pm$1.8 & $-$16.3$\pm$1.7 & \\
             2 & $-$17.4$\pm$1.8 & $-$14.3$\pm$2.0 & $-$14.1$\pm$1.9 & $-$15.4$\pm$2.1 & \\
             3 & $-$13.8$\pm$1.7 & $-$10.8$\pm$1.8 & $-$10.9$\pm$1.7 & $-$11.9$\pm$1.9 & \\
             4 & $-$16.1$\pm$2.1 & $-$12.5$\pm$2.0 & $-$11.8$\pm$1.9 & $-$13.3$\pm$2.6 & $-$14.4$\pm$2.2 \\
\hline
\end{tabular}
\end{center}
\end{table*}

Now we use the derived parameters to estimate stellar radii for all 5 objects.  Following the same practice as \citet{Herrero1992}, the radius $R$ is defined by
\begin{equation}
  5\log\frac{R}{R_{\odot}} = 29.57 - (M_V - F_m)
\end{equation}
where $F_m$ is the logarithm of the integral of the model stellar flux within the $V$ passband.  Finally, the stellar mass, luminosity, mass loss rate $\dot{M}$ and modifed wind momentum ($D_{\rm mom}$) follow from the values derived for surface gravity and $R$ (table~\ref{tab:masses}). $D_{\rm mom}$ is obtained from the modified wind-momentum luminosity relationship \citep[WLR, see][]{puls96}:
\begin{equation}
 \log D_{\rm mom}=  \log (\dot{M}v_\infty R^{0.5})= x \log L/L_\odot + D_{\rm 0}
 \end{equation}
where values for $x$ and $D_{\rm 0}$ can be found for example in \citet{mokiem07}. For the wind terminal velocity ($v_\infty$), that we cannot derive from our spectra, we adopt the canonical relationship between escape velocity $v_{\rm esc}$ and $v_\infty$, that $v_{\rm esc}/v_\infty$= 2.65, \citet{KP00}, (but see \citet{garcia14} for the limitations of this assumption).  On this basis we find all targets are consistent with the expectations from the WLR, indicating normal stellar winds for the spectral types.

  The derived masses for \#1338, \#1273, MSP 182 and MSP 199 are broadly consistent with the 40--60~M$_{\odot}$ evolutionary tracks computed by \citet{Ekstrom2012}, taking stellar rotation into account.  For the error propagation we have considered a somewhat large error for $M_{\rm V}$ ($\sim$ 0.3 mag) but do not take into account an error in the adopted distance.   The high mass estimated for MSP 183 is subject to the greatest uncertainty, thanks especially to the less well-validated visual extinction (a change of 0.1 in $A_V$ alone propagates directly to a 10\% change in stellar mass -- the error could easily be twice this).

\subsection{Radial velocity analysis}
\label{sec:rvs}

Our approach to measuring the radial velocities of the two O stars, relative to the reference objects, has been to use cross-correlation in the blue spectrum from 360 nm to 510 nm. Tapering to eliminate end effects reduces the effective wavelength range to 367 -- 503 nm.  Routines from two independent astronomical packages (DIPSO and IRAF) have been applied to this task, thereby testing for differences in numerical handling, such as in the method of spectrum normalisation.  In both cases, the parts of the spectrum containing the stronger diffuse interstellar bands at $\lambda\lambda$s 442.7 nm and 488.2 nm were removed and replaced by linear interpolations.  Each epoch of observation has been cross-correlated with each reference object, yielding a grid of $3\times3$ and $4\times3$ RV measurements for \#1338  and \#1273, respectively.  The best agreement, to well within the mutual errors, between the two independent measures was achieved when the fits included only the top half of the main cross-correlation function (CCF) peak.  We give the results based on the DIPSO routines as these are accompanied by explicit errors on the individual CCF peak fits (see Table~\ref{tab:rvmeasures}).  They are specified in km~s$^{-1}$ on a scale that sets the mean heliocentric velocity of MSP 182 \citep[30.5 km s$^{-1}$][]{Rauw2011} to zero.       

In extracting these results we need to assume that the mean RVs obtained by \citet{Rauw2011} for the reference objects remain sound. It is reassuring that there is no {\em strong} systematic effect apparent that undermines this -- but there is some sign in Table~\ref{tab:rvmeasures} of a $\sim$3 km s$^{-1}$ offset between MSP 182 on the one hand, and MSP 199 and 183 on the other.  This contrast may have its origin in the spectral type difference \citep[section~\ref{sec:parms} and][]{Rauw2011} and could argue for weighting the epoch averages in favour of cross-correlations with MSP 182 since this star most closely resembles \#1273 and \#1338.  Rather than introduce arbitrarily-chosen unequal weights in forming the means, we just take note that the derived means potentially underestimate the blueshifts by up to $\sim$3 km s$^{-1}$.

We have also carried out fits of the HeII~$\lambda$541.1 nm line profile alone as a further comparison. The overall means obtained by this route for the two targets are $-31.1 \pm 1.7$ and $-17.5 \pm 2.2$ km s$^{-1}$.  To within the errors, these outcomes are consistent with the cross-correlation results but have the disadvantage of a systematic dependence on the choice of continuum around the one line, and of some wind effect on the profile.  Accordingly, we prefer the tabulated cross-correlation measures that capitalise on several spectral features. These In Table measures are $-29.4 \pm 1.7$ and $-14.4 \pm 2.2$ for \#1338 and \#1273 respectively (Table~\ref{tab:rvmeasures}).

The epoch to epoch radial-velocity variation of each star is compatible with measurement error, with a little more variation apparent in \#1273.  The consistency and the pattern from the observations, spanning 10 and 17 days for \#1338 and \#1273 respectively, indicate that binary motion is not prominent in either target and that the measured radial velocities are likely to be a good guide to the systemic motions of our two targets. We have set up Monte Carlo simulations to quantify the probabilities involved, using the same method as \citet{Rauw2011}: the reader is referred to section 4, figure 14 and tables 7 and 8 of this previous study for the details.  Like \citet{Rauw2011}, we: consider only periods up to 100 days; assume zero eccentricity at periods shorter than 4 days; adopt flat distributions in both eccentricity ($0 < e < 0.9$) and binary mass ratio ($0.1 < q < 1$). For \#1338, the maximum radial velocity change over the 3 epochs of data is under 5 km~s$^{-1}$.  In this case, for a plausible stellar mass of 50 M$_{\odot}$, so small a variation all but rules out an orbital period of under a month -- our simulation indicates under 0.5\% of binaries could produce such a signature.  The 'missed binary' percentage rises to $\sim$2\% and $\sim$7\% of binaries for periods of respectively 1-2 months, and $>$2 months.  The analogous probabilities for \#1273, observed on 4 dates, and yielding a maximum radial velocity change less than 7 km s$^{-1}$, would be very similar, i.e. no more than 0.2\%, 1.5\% and 6\%.   

It thus seems most likely that \#1338 and \#1273 can be regarded, for present purposes, as single stars whose spectra provide reliable radial velocities.  
%The same is very likely to be true of \#1273, although it may yet prove to be the primary of a relatively long period system.  
%In both cases, our working hypothesis is that the mean measured radial velocity in table~\ref{tab:rvmeasures} adequately represents the systemic velocity.

\section{Further considerations -- a preliminary discussion}
\label{sec:appraisal}

Before going onto the relative proper motion data for the entire sample, some comment is appropriate on three issues: how the radial velocities of \#1273 and \#1338 fit into what is already known about the Wd2 sightline kinematics; what can be gleaned from the literature on potential ejections from Wd2 so that we do not overlook them; other evidence regarding some sample members that already implies no association with Wd2 or else that they are not ejections.

\subsection{On the measured radial velocities of \#1273 and \#1338}
\label{sec:rv-implications}

Both of the O stars followed up with X-shooter observations have been shown to have significantly blueshifted radial velocities relative to the previous measurements of Wd2 by \citet{Rauw2011}.  This earlier study noted that most of the cluster stars were compatible with systemic radial velocities in the range 20--30 km s$^{-1}$, in the heliocentric scale.  In the \citet{Rauw2011} sample, four O stars were found to show no or negligible binary motion, and the mean radial velocity among these is close to $+$27 km s$^{-1}$.  Three of this group of four are our reference objects, MSP 182, 199 and 183.  Referred to MSP 182, the measured relative radial velocities of our two targets are $-29.4\pm1.7$ and $-14.4\pm2.2$ km s$^{-1}$ (Table~\ref{tab:rvmeasures}): accepting 30.5 km s$^{-1}$ as the mean heliocentric radial velocity of MSP 182 \citep{Rauw2011}, these convert to absolute heliocentric values of $+1$ and $+16$ km s$^{-1}$ (hereafter rounding to the nearest km s$^{-1}$).  Expressed relative to the Wd2 cluster mean taken as $+$27 km s$^{-1}$, the radial velocities of \#1338 and \#1273 become $v_r = -26$ and $-11$ km s$^{-1}$, similarly rounded.

Already, the first of our targets, \#1338, comes close to fitting the commonly-understood definition of a runaway star, with a space velocity exceeding 25 or 30 km s$^{-1}$ (depending on preferred definition).  Later, account will be taken of the transverse motion ({see Section~\ref{sec:gaia}).  %Should the time since ejection be as long as 1.5~Myr, the transverse speed needed to achieve the observed angular separation at a distance of 5 kpc becomes 20 km s$^{-1}$.  This combines with the measured blueshift of $\sim$ 26 km s$^{-1}$ to yield a space velocity of 33 km s$^{-1}$.  There is considerable uncertainty in the age of Wd2: the most recent and exhaustive cluster isochrone fits by \citet{Zeidler2017} are compatible with an age range from 0.5 to 2 Myr.  Even reducing the distance to 4 kpc, the shortest likely, it would be hard to get the space velocity down below 30 km~s$^{-1}$.  In short, kinematically, object \#1338 is well set up as a potential ejection from Wd2.  We complete the picture with a proper motion in Section~\ref{sec:gaia}.  

For object \#1273, the circumstances are not so clear-cut.  To expose this, it is helpful to shift from the heliocentric frame to the Local Standard of Rest (LSR).  The correction from heliocentric to LSR is $-9.0$ km~s$^{-1}$ \citep[using the solar motion data from][]{Schoenrich2010}.  This places Wd2 at a $v_{LSR} = 18$ km~s$^{-1}$ and object \#1273 at $v_{LSR} = 7$ km~s$^{-1}$.  CO observations of the region have revealed molecular clouds at $v_{LSR} = 16$, $4$ and $-4$ km s$^{-1}$ \citep{Furukawa2009}, that all show elevated temperatures consistent with heating by Wd2 and its HII region RCW~49 \citep{Ohama2010}.  \citet{Dame2007} presented absorption measurements showing that the 4 and $-$4 km~s$^{-1}$ clouds are in front of the 16 km~s$^{-1}$ cloud, and argued that Wd2 is in a cavity in front of the latter, whilst behind the former.   Given the uncertainty in the measured stellar radial velocity, and the location of \#1273 on the edge of the $+$4 km s$^{-1}$ cloud, the evidence may be read as a potential kinematic relation between the two.  %This does not close off the possibility that \#1273 is also an ejection from Wd2.
%Comparing the position of \#1273 with the CO maps presented by \citet{Furukawa2009}, it can be seen that it lies about 2 arcmin away from the lowest CO contour plotted for the $+4$ km~s$^{-1}$ cloud (see figure 1 in the cited study).

%Object \#1273 may yet be the result of an ejection from Wd2, as seems likely for object \#1338.  In this case, the full 3-D space velocity compatible with its angular separation of 11.3 arcmin from the centre of Wd2 and its measured blueshift would be 16 km~s$^{-1}$ (again adopting a time of flight of 1.5~Myr, and a distance of 5~kpc).  Viewed within a framework that imposes a binary choice between 'cluster members' and 'runaway stars', this object fails to fit easily into either category.

\subsection{Candidate massive star ejections from Wd2 from prior literature}
\label{sec:literature}

%The implications of our X-shooter data are that \#1338 is a good candidate for what would be a fairly low-velocity runaway, while the kinematic status of \#1273 not yet clear.

The work by \citet{Rauw2011} pointed to two more early O stars with, again, significantly negative radial velocities relative to the other Wd2 members.  These were MSP 18 and 171, for which the mean heliocentric radial velocities obtained were $(-1.1 \pm 1.8)$ and $(-9.3 \pm 6.9)$ km~s$^{-1}$.  Relative to the mean for the cluster these become $v_r = -28$ and $-36$ km s$^{-1}$ -- similar to, if a little larger than, our result for \#1338.  Here, if these stars are ejections, the transverse velocities should not contribute very much to the total space velocity since both objects are only modestly displaced from the bright cluster core.  The repeat observations obtained by \citet{Rauw2011} did not reveal obvious binary motion in either case.
%The MS-II estimates for the extinction towards MSP 18 and MSP 171 are $A_0 = 5.93$ and $6.51$, to be compared with $6.17$ and $6.67$ for \#1338 and \#1273 (see Table~\ref{tab:list}).

We mentioned already in Section~\ref{sec:hinterland} the proposal by \citet{RomanLopes2011} that the Wolf-Rayet (WR) stars WR20c and WR 20aa, both typed as O2If$^*$/WN6, were ejected from Wd2.  Their angular separations from Wd2 bracket that of object \#1338, in being $\sim$25 and $\sim$15 arcmin respectively. \citet{RomanLopes2011} were particularly struck by the fact that the line on the sky joining WR 20c and WR 20aa passes through the stellar-density centre of Wd2.  Whilst the arresting sky geometry might suggest ejection, there is no back up from radial velocities since these are very hard to measure from spectra dominated by strong and broad WR line emission.

A challenge presented by WR 20c, is that its extinction is appreciably higher than prevails in and around Wd2 itself.  \citet{RomanLopes2011} provide a measurement of the colour excess, $E(B-V) = 2.9$, which is close to the value of 2.8 consistent with the MS-II best fit values, $A_0 \simeq 10.3$ and $R_V \simeq 3.7$ (see table~\ref{tab:list}).  The visual extinction to this object, although approximate, lies well above and outside the range $5.5 < A_0 < 7.5$ we have used here to select O stars more likely to have a physical association with Wd2.  If this WR star is an ejection from Wd2, it has moved behind a very substantial, localised column of gas and dust - representing around two-thirds of the column, in addition, that accumulates towards Wd2 itself (assuming no great change in dust grain properties).  In this circumstance, the heating and ionizing effect of WR 20c on a nearby dark cloud. might well be noticeable, and yet there is no clear sign of heated dust in WISE data.  This absence along with the enormous additional dust column raises the suspicion WR 20c may be a background object.

For WR 20aa, the extinction is not obviously problematic -- the MS-II measures are $A_0 \simeq 5.6$ and $R_V \simeq 4.3$, implying $E(B-V) = 1.3$ \citep[cf. 1.5 from][]{RomanLopes2011}. This is less than for Wd2 itself, but by no more than a magnitude.

At this point it is appropriate to also mention the brighter WR star, WR 21a, located $\sim$16 arcmin away from the centre of Wd2.  It is too bright to have been included in the MS-II catalogue.  This star is known to have an O star companion and the binary orbit has been analysed \citep{Tramper2016}, but there is no reported measurement of a systemic radial velocity. \cite{RomanLopes2011} drew attention to it as lying on a vector almost perpendicular to the line on the sky joining WR 20c and WR 20aa, and wondered if it too is an ejection.

%At this point in the census of candidate O-star ejections, we have 3 likely candidates (MSP 18, 171 and \#1338), an ambiguous example (\#1273) and the less compelling case of WR 20c.

%Next, we return to the spread of OB candidates around Wd2 revealed in the data from MS-II (see Figure~\ref{fig:map}).  Our main O-star sample (Table~\ref{tab:list}) contains 19 objects.  Both the X-shooter targets are in this set, along with WR 20aa.  Of the remaining 16 we have spectroscopic effective temperature estimates for 6.  The hottest of the 6 at $\sim$42 kK, object \#1236, is only just outside the cut at 10 arcmin radius from the cluster centre: we pair it with the also very hot \#1273 that is at a similar angular separation, directly on the other side of Wd2.  %For now, we categorise it as a marginal case, like \#1273.  
%The other 5 O stars spread in temperature from $\sim$39~kK (O5.5-6) up to $\sim$35 kK (O7.5), and they are all main sequence stars based on the available surface gravity estimates.  This leaves 10 with only photometric constraints on stellar parameters.  MS-II showed that all are plausibly at least early B stars, and that 7 of them  are likely to be O stars. 

%\subsubsection{Are any of the candidate objects in clusterings?}
%\label{clusterings}
\subsection{A distinct OB grouping and other 'unrelated' objects}
\label{sec:unrelated}

\begin{figure}
\begin{center}
\includegraphics[width=0.9\columnwidth]{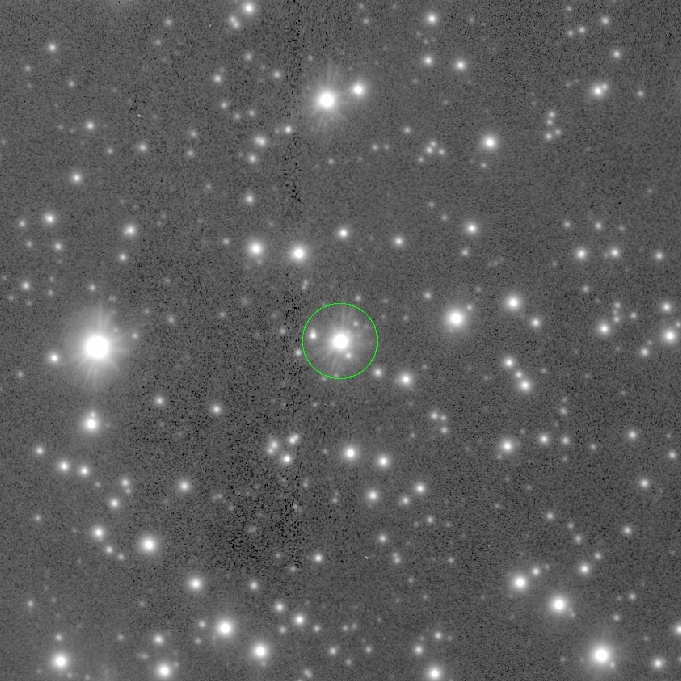}
\includegraphics[width=0.9\columnwidth]{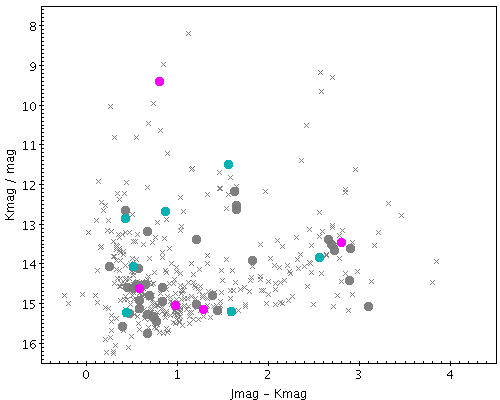}
\caption{
Top panel: a $3\times3$ arcmin$^2$ cut-out VPHAS$+$ $i$-band image around object \#1338.  The stellar density in the area is significant, but there is no sign of an over-density of fainter stars around \#1338.  The green circle has a radius of 10 arcsec.  Lower: the 2MASS $K$ versus $J - K$ colour-magnitude diagram, around \#1338.  All objects within a radius of 3 arcmin are included, and they are coloured according to how far they are from \#1338: bright pink implies $<0.25$ arcmin separation; cyan $<0.5$ arcmin; grey (circles) $<1$ arcmin.  Star \#1338 itself is in pink, located at $K \simeq 9.4$.)
}
\label{fig:1338-cut-out}
\end{center}
\end{figure}

\begin{figure}
\begin{center}
\includegraphics[width=0.9\columnwidth]{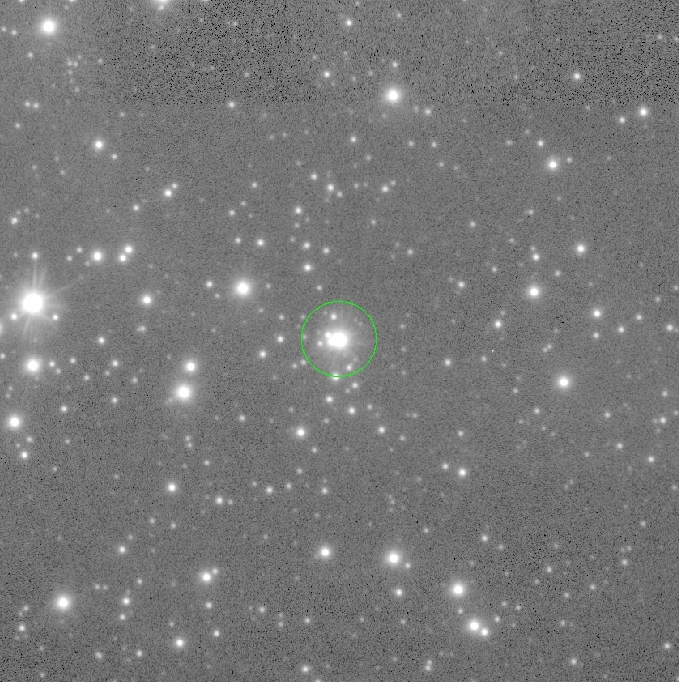}
\includegraphics[width=0.9\columnwidth]{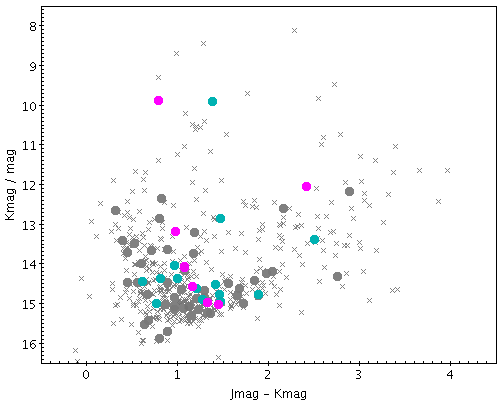}
\caption{
The same as Figure~\ref{fig:1338-cut-out} for \#1567.  In this case there is sign of a rise in fainter-star stellar density around \#1567, that is backed up by evidence in the NIR CMD that there are the beginnings of a credible localised cluster main sequence (the line of 7 bright pink points, with \#1567 at the top -- two of the stars have almost identical $J-K$, $K$ magnitudes.)}
\label{fig:1567-cut-out}
\end{center}
\end{figure}

We now pose the question as to which of the O star candidates in Table~\ref{tab:list} exhibit properties that argue against a connection with Wd2.  One such property -- already used against WR 20c -- is appreciably higher extinction that could point to greater distance.  We deal with this first in considering the candidate objects also included in the lower part of Table~\ref{tab:list}.

Above $A_0 = 7.5$, the MS-II catalogue lists only 4 good OB stars located within 10 arcmin of the centre of Wd2 - to be compared with 36 in the range, $5.5 < A_0 < 7.5$.  Just one of the four is close to the dividing line with its extinction given as 7.74.  It was this dramatic dropping away in the extinction distribution that prompted the choice of $A_0 = 7.5$ as a selection boundary and the division into two of Table~\ref{tab:list}. All the same, it is arbitrary, and we now examine it in more detail.

A notable feature of the sky distribution of the $A_0 > 7.5$ objects, more than 10-arcmin radius from Wd2, is that 14  (out of 21 altogether) appear loosely grouped around $\ell = 283.9^{\circ}$, $b = -0.9^{\circ}$ (see Figure~\ref{fig:map}).  The extinctions for these stars fall within the range $7.5 < A_0 < 8.6$ (Table~\ref{tab:list}), hinting this is not necessarily an asterism. 8 of the 14 objects form a dense inner cluster.  Perhaps this is an association, independent of Wd2, at a different, possibly greater, distance.  These stars are identified in Table~\ref{tab:list} by the comment '$A_0 \sim 8$ BG' -- BG is short for 'background group'.  We add to the group two stars, \#0986 and \#0987, with $A_0 = 7.33$ and 7.40 that belong otherwise to the upper part of Table~\ref{tab:list}.  Their location in the same part of the region and their almost-as-high extinction indicate this could be appropriate. 

The $K$ magnitudes of the proposed grouping are fainter, ranging from 9.6 down to 12.7.  Presently, none of these stars has a spectrum, and so only the photometric estimates of effective temperature are available: they run from $\sim30$ up to $\sim40$~kK.  Of the 6 remaining higher-extinction objects (excluding WR~20c), 2 are in the same sky area as the $A_0 \sim 8$ group, while the other 5 are widely scattered (see Figure~\ref{fig:map}). 

A separate question to ask of the candidate objects in the upper part of Table~\ref{tab:list} is whether any of them seems likely to have an associated cluster of fainter stars.  Any that do cannot have formed in and have been ejected from Wd2 -- ejected stars do not take clusters with them.  We have examined VPHAS+ $i$ band images for relative stellar over-densities around these objects, and have constructed 2MASS colour-magnitude diagrams (CMDs) to look for evidence of coherent cluster sequences.  VPHAS+ $i$ offers the higher dynamic range in that most of the 19 stars have $i$ magnitudes between 12 and 13, easily permitting a well-defined search for companion objects that are 6--7 magnitudes fainter, reaching down into the later A spectral-type range (assuming $M_i \sim -4$ for late O stars, and $M_i \sim 2.5$ for late A stars).  From 2MASS, we have constructed $K$ versus $J-K$ diagrams that optimise for minimum impact of extinction on the vertical axis, and a better dynamic range in the colour dimension on the horizontal axis. 
%But we consider both, since the impact of extinction variation will be higher in $i$.

The results of this exercise are illustrated in Figures~\ref{fig:1338-cut-out} and \ref{fig:1567-cut-out}, which include an $i$-band cut-out and 2MASS CMD for respectively \#1338 as an example of a star without any evident clustering around it, along with the same for \#1567 which emerges as the only object colocated with a potential cluster.  The $i$ over-density in this second case within 0.5 arcmin of the star is 1.9$\sigma$ (assuming Poisson statistics), and it is the only example that indicates a convergence onto a nearly vertical cluster main sequence as the area of 2MASS selection around the star is shrunk down onto the star.  The next best example is \#1102 for which the over-density significance is down to 1.4$\sigma$, and yet the NIR CMD is unconvincing.  Consequently, only \#1567 is proposed with any confidence as associated with its own cluster, thus becoming an unlikely ejection.

%\subsubsection{A final narrowing down}
%\label{sec:final-reduction}
Last of all, some words on object \#0708 are appropriate. The photometric estimate of this object's effective temperature (in Table~\ref{tab:list}, part (a)) suggests it is a B1--O7 star.  It can be seen in Figure~\ref{fig:map} that this star is positioned close to the bow-shaped nebulosity near $\ell = 283.5^{\circ}$, $b = -1.0^{\circ}$.  The nebula here is NGC 3199, an HII region excited by WR 18, a bright Wolf-Rayet star, likely to be $\sim$2 kpc away \citep{WRcat2001}.  The angular separation between \#0708 and WR 18 is just 1.14 arcminutes, and the 2MASS K magnitude of \#0708, the brightest in our selection, is only 0.26 fainter than that for WR 18.  Accordingly a physical association between WR 18, NGC 3199 and \#0708 at $\sim$2 kpc appears to be the better bet.  In addition, recent work by \citet{toala2017} challenges an older view that WR 18 is a runaway from Wd2: they do this on grounds of proper motion data on the WR star and stars near it, and also the far-infrared morphology of NGC 3199.

\section{Relative proper motions from Gaia DR2}
\label{sec:gaia}

Whilst Gaia DR2 parallaxes are not yet good enough for estimating the distance to Wd2 or the potentially associated O stars, the proper motion (PM) data are already very useful.  In and around Wd2 they are typically 5--7 mas yr$^{-1}$ in magnitude, and we can propagate the uncertainties in PMs relative to the cluster mean from those stated in the released database.  The first step in measuring the relative PMs is to determine the Wd2 cluster mean.

The Gaia DR2 database was cross-matched with the objects in Figure~\ref{fig:members}, after the stars lying outside the core $8 < K < 12$ range had been trimmed off.  The remaining objects were then checked for astrometric quality: one object was removed for being flagged as 'duplicate', while some were rejected on grounds of high excess astrometric source noise \citep[using $\epsilon_i$, see the $G$-dependent thresholds in table B.1 of][]{Lindegren2018}.  We also excluded MSP 18 and MSP 171, since these are candidate runaway stars (see Section~\ref{sec:literature}). This left us with 25 objects.  The median proper motion derived from them is $\mu_{\alpha,*} = -5.172$ mas~yr$^{-1}$, $\mu_{\delta} = 2.990$ mas~yr$^{-1}$.  Further reduction of the sample to limit it to 18 stars in the densest part of the cluster, occupying a little over one square arcminute, only altered the last decimal place in these measures.  The sample standard deviations (for the 25 stars), created by a combination of astrometry error and velocity dispersion with the cluster, are 0.204 and 0.164 mas yr$^{-1}$ in RA$\ast$ and Dec respectively.   

Converting the median cluster PM into Galactic co-ordinates, we obtain $\mu_{\ell,*} = -5.970$ mas yr$^{-1}$, $\mu_{b} = -0.227$ mas yr$^{-1}$.  It is encouraging and to be expected that the representative cluster PM emerges as almost entirely in the longitude direction: a cluster as young as Wd2 should closely follow Galactic disc rotation.  These values have been subtracted from the proper motions, in Galactic coordinates, of all the objects listed in Table~\ref{tab:list} (see the Appendix for the names of Gaia DR2 sources cross-matched to them). The resulting relative PMs, and their errors, are visualised in Figure~\ref{fig:pm-map}.

%To compare with this, we have computed the expected $\mu_{\ell,*}$, for a range of distances, applying the following specification: we adopt a Galactic disc circular speed, $\Theta_0$, of 240 km s$^{-1}$ a distance from the Sun to the Galactic centre, $R_0 = 8.34$ kpc, and use $(\Theta_0 + v_{\odot})$ = 255.2 kms s$^{-1}$ \citep[parameters from][]{Reid2014}; we correct for the Sun's motion with respect to the Local Standard of Rest using $u_{\odot}$ = 11.1 km s$^{-1}$ \citep{Schoenrich2010}.  In this manner, we arrive at predictions for $\mu_{\ell,*}$ of between $-$6.83 and $-$5.98 mas yr$^{-1}$ for distances to Wd2 ranging from 4 to 7~kpc -- certainly higher than observed, at the shorter distances. %We can bring the matching distance down by tweaking to e.g. $V_0 = 220$ km~s$^{-1}$ and $R_0 = 8.3$~kpc: these values would permit the median cluster PM to be matched to within 
%The Gaia DR2 0.1 mas systematic error translates to roughly 300 pc.  A random source of error is the peculiar motion relative to mean rotation exhibited by regions of high mass star formation: \cite{Reid2014} found the magnitude of this to be up to $\sim$10 km s$^{-1}$.  This introduces a further $\pm$0.15 mas yr$^{-1}$ or a distance uncertainty in the region of 0.5 kpc. 

\begin{figure*}
\begin{center}
\includegraphics[width=1.55\columnwidth]{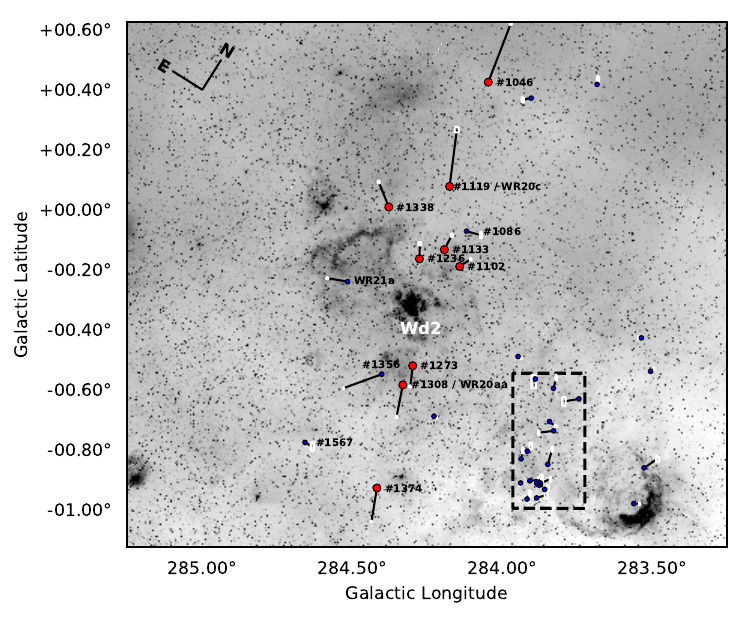}
\caption{
The relative proper motions derived for all objects in Table~\ref{tab:list}, excepting \#1550 for which the Gaia DR2 data are problematic.  MSP 18 and MSP 171, near the cluster centre are not included either since most of their space motion is in their radial velocities.  The direction and amount of proper motion is represented by the black line drawn away from each object: the length of each vector corresponds to the angle each object would move through in 0.2 Myr.  The red circles pick out the stars listed in Table~\ref{tab:gaia} (excepting WR 21a -- see text), whose relative PM vectors point outward from Wd2.  The remaining objects and WR 21a, are shown as blue dots, and in many cases the relative proper motion is so small that the PM vector is hard to see.  The white box at the end of each relative PM vector indicates the directional error zone propagated from the Gaia DR2 uncertainties.
}
\label{fig:pm-map}
\end{center}
\end{figure*}

\begin{table*}
\caption{Proper motions and related quantities for stars potentially qualifying as runaways from Wd2.  The Gaia DR2 proper motions appear in columns 2 and 3.  Columns 4 and 5 give the proper motion relative to the WD2 cluster median, re-expressed in Galactic coordinates.  %Column 6 lists the difference in angle between the relative proper motion and the line of flight from the Wd2 reference position at $\ell = 284^{\circ}.27$, $b = -0^{\circ}.334$. 
Column 6 is the (distance-independent) travel time from the fiducial position, $\ell = 284^{\circ}.27$, $b = -0^{\circ}.334$.
Column 7 gives $v_t$ for the working distance of 5 kpc, accompanied by the range allowing for varying the distance from 4 to 6 kpc in brackets and taking into account random error in the total relative proper motion.  The last column gives the full space velocity, $v_s$, where known -- again giving the range for $4 \leq D \leq 6 $ kpc in brackets.   The 3 objects below the horizontal line are additional to the MS-II sample of stars in Table~\ref{tab:list}.}
{\centering
\begin{tabular}{lccrrcll}
\hline  
Object & \multicolumn{2}{c}{Proper motion} & \multicolumn{2}{c}{Relative PM} & Travel time & $v_t$ & $v_s$ \\
    & \multicolumn{2}{c}{$\mu_{\alpha,*}$, $\mu_{\delta}$ mas/yr} & \multicolumn{2}{c}{$\Delta\mu_{\ell,*}, \Delta\mu_{b}$ mas/yr} & Myr & km/s & km/s \\
\hline
Wd2 centre & -5.173$\pm$0.041 & 2.995$\pm$0.033 & $-$ & $-$ &  &  &  \\
\hline
%\#0673 & $-$5.229$\pm$0.050 & 3.222$\pm$0.048 & $-$0.198$\pm$0.030 & 0.177$\pm$0.088 & {\bf 57} & -- & 6 & \\
%\#0693 & $-$5.007$\pm$0.056 & 2.885$\pm$0.050 & 0.170$\pm$0.032 & 0.010$\pm$0.091 & {\bf 177} & -- & 4 & \\
%\#0893 & $-$5.637$\pm$0.071 & 2.964$\pm$0.065 & $-$0.405$\pm$0.036 & $-$0.259$\pm$0.108 & 8 & {\bf 4.2} & {\bf 11} & \\
%\#0994 & $-$5.379$\pm$0.029 & 2.984$\pm$0.029 & $-$0.197$\pm$0.023 & $-$0.104$\pm$0.069 & 3 & {\bf 5.7} & {\bf 5} & \\
\#1046 & $-$4.441$\pm$0.051 & 6.675$\pm$0.051 & $-$1.344$\pm$0.062 & 3.494$\pm$0.064 & 0.76 & 89 (69--110) & \\
\#1102 & $-$5.489$\pm$0.051 & 3.722$\pm$0.054 & $-$0.653$\pm$0.062 & 0.440$\pm$0.067 & 0.88 & 19 (15--23)  & \\
\#1133 & $-$5.092$\pm$0.067 & 3.967$\pm$0.074 & $-$0.449$\pm$0.071 & 0.859$\pm$0.088 & 0.80 & 23 (18--29) & \\
%\#1164 & $-$5.334$\pm$0.044 & 3.006$\pm$0.046 & $-$0.171$\pm$0.022 & $-$0.062$\pm$0.084 & {\bf 63} & -- & 4 & \\
\#1236 & $-$4.707$\pm$0.075 & 3.781$\pm$0.079 & $-$0.024$\pm$0.080 & 0.907$\pm$0.090 & 0.68 & 22 (17--27) & \\
\#1273 & $-$5.716$\pm$0.060 & 1.872$\pm$0.056 & 0.143$\pm$0.071 & $-$1.244$\pm$0.066 & 0.54 & 30 (23--37) & 32 (26--38) \\
\#1338 & $-$3.856$\pm$0.064 & 3.949$\pm$0.058 & 0.605$\pm$0.068 & 1.504$\pm$0.075 & 0.80 & 39 (30--48) & 47 (40--54) \\
%\#1356 & $-$3.675$\pm$0.035 & 1.047$\pm$0.033 & 2.276$\pm$0.024 & $-$0.831$\pm$0.073 & {\bf 37} & -- & 58 & \\
\#1374 & $-$5.973$\pm$0.037 & 1.126$\pm$0.033 & 0.324$\pm$0.053 & $-$2.011$\pm$0.049 & 1.08 & 48 (38--60) & \\
WR20aa & $-$5.876$\pm$0.055 & 1.186$\pm$0.054 & 0.374$\pm$0.065 & $-$1.908$\pm$0.067 & 0.48 & 46 (36--57) & \\ 
WR20c  & $-$3.715$\pm$0.154 & 6.119$\pm$0.140 & $-$0.434$\pm$0.136 & 3.412$\pm$0.165 & 0.44 & 82 (60--106) & \\
\hline
MSP 18 & $-$5.295$\pm$0.085 & 3.292$\pm$0.085 & $-$0.260$\pm$0.085 & 0.273$\pm$0.100 & $\sim$0.1 & 9 (6--12) & 29 (28--31) \\
MSP 171 & $-$4.761$\pm$0.063 & 2.898$\pm$0.059 & 0.402$\pm$0.072 & 0.133$\pm$0.070 & $\sim$0.1 & 10 (7--13) & 37 (34--42) \\
WR 21a & $-$4.025$\pm$0.051 & 2.518$\pm$0.045 & 1.226$\pm$0.062 & 0.205$\pm$0.059 & 0.65 & 30 (23--37) & \\
\hline

%\#1550 & $-$5.336$\pm$1.203 & $-$9.452$\pm$1.249 & {\sl $-$0.18} & {\sl $-$12.45} & {\sl 28} & \\
%\hline
%\#1567 & $-$5.656$\pm$0.192 & 2.991$\pm$0.179 & $-$0.435 & $-$0.246 & {\bf 78} & & 12 & N \\
\hline
\end{tabular}
}
\label{tab:gaia}
\end{table*}

Data on the relative PMs of those stars revealed in Figure~\ref{fig:pm-map} as potential ejections, are set out in Table~\ref{tab:gaia} along with the main derived quantities: we specify the relative PM and error, along with estimates for: the travel time as given by ratio of the relative PM to the angular separation from Wd2 (this assumes constant ejection speed); the transverse speed, $v_t$ in the plane of the sky; finally, where we have it, $v_s = \sqrt{v_t^2+v_r^2}$, the full space motion.  The error budget does not include the $\sim$0.1 mas basement error advised by \cite{Lindegren2018} since this is systematic and unlikely to influence relative motions measured within $\sim$1 degree on the sky.  The PM errors from the individual-source random errors and correlations are indicated in Figure~\ref{fig:pm-map}.  It is important and of course very useful that these errors are mostly small.  We did not compute the relative PM for one object -- \#1550: this had to be left out because the Gaia DR2 catalogue flags it as duplicated, and indeed the stated errors are very large. In all other cases, the available astrometry is based on at least 15 distinct epochs and all, bar WR 20c, have $u = \sqrt{\chi^2_\nu}$ safely below the cut recommended by \cite{Lindegren2018} (see their equation C.1).  For WR 20c -- by far the reddest object in the sample -- $u$ is $\sim$9, rather than $\lesssim 3$ as recommended by the $G$-sensitive cut.

The most striking features of the derived relative PMs are the number of likely ejections (up to 12, at first sight) and their apparent on-sky alignment with a direction almost perpendicular to the Galactic equatorial plane.  Both WR 20c and WR 20aa are included in this group, with WR 20c as the only example from the lower half of Table~\ref{tab:list}.  In general terms, the expectation that stars with similar extinctions to the stars in Wd2 would more likely be associated with the cluster has been borne out: seven objects from the upper half of Table~\ref{tab:list} have relative PMs compatible with ejection from Wd2.  This group includes both \#1338 and \#1273.  The earliest ejection appears to be \#1374, for whom the travel-time estimate is a little over 1 million years.  

There is one exception to the general alignment: the binary, WR 21a also has a significant PM relative to Wd2, but it stands apart in being directed almost at right angles to the main ejection axis (see Figure~\ref{fig:pm-map}).  As for WR 20c and WR20aa, no systemic radial velocities are yet available (and will be challenging to obtain).  But there is an important point of difference in that the raw PM in this case, $\mu_{\ell,*} = -4.744\pm0.048$, $\mu_b = -0.022\pm0.048$ mas yr$^{-1}$, has a negligible Galactic latitude component.  This offers an alternative reason for the difference in its PM compared with Wd2: WR 21a may simply be unrelated and at greater distance than the cluster.

We include in Table~\ref{tab:gaia} the relative PMs of MSPs 18 and 171, the objects measured by \cite{Rauw2011} to have significant blueshifted radial velocities (see Section~\ref{sec:literature}).  At amounts corresponding to transverse velocities of no more than $\sim$10 km s$^{-1}$ at 5 kpc, in both cases, it is clear that most of their motion is indeed radial (see section~\ref{sec:literature} and Table~\ref{tab:gaia}).  Their relative PMs are directed away from the main clustering, although defining this precisely is rendered impossible by both the greater significance of the errors and proximity to the central region.  However, the existence of some proper motion and the positioning of both toward the periphery of the main Wd2 cluster allows a very rough estimate of the elapsed time since expulsion: it cannot be much more than 10$^5$ years in either case. 

A further outcome is support for the presence of a distinct, potentially background, OB-star grouping (discussed above in Section~\ref{sec:unrelated}). There are evidently small and disordered relative PMs within or near the rectangular region picked out in Figure~\ref{fig:pm-map}, no star in this region presents with $v_t$ approaching 20 km s$^{-1}$ (at 5 kpc), that is also directed away from Wd2.   

There is one prominent example of an object with a large relative proper motion that does not trace back to the vicinity of Wd2: this is object \#1356, below Wd2 in Figure~\ref{fig:pm-map} moving at $\sim$60 km s$^{-1}$ relative to Wd2 (if 5~kpc away).  It has a significant absolute proper motion away from the Galactic Plane (of $-1.078\pm0.047$ mas yr$^{-1}$) suggesting it could be an escape, but from where is as yet unidentified.  Object \#1086 above Wd2 in Figure~\ref{fig:pm-map} is a more modest example: it is much more highly extinguished ($A_0 = 9.7$) and perhaps well beyond Wd2.

Finally, we note that \#1567, the star picked out as most likely to be in a cluster of its own (Section~\ref{sec:unrelated}), is seen in Figure~\ref{fig:pm-map} to show little relative proper motion. %This is as expected. 

\section{Discussion}
\label{sec:discussion}

\citet{Fujii2011, Fujii2012} carried out simulations of massive young clusters, citing Wd2 as such a cluster, in order to quantify the likely ejection yield from early-phase intra-cluster dynamical interactions.  Essentially, the simulated ejections are all the product of interaction with a steadily hardening binary in the cluster core.  \citep{Fujii2011} predicted that clusters like Wd2 should eject $5.2 \pm 1.6$ stars more massive than 8 $M_{\odot}$ in their first Myr.  Their results also showed that the distribution of space velocities is skewed towards the low end, heavily favouring the 20--30 km~s$^{-1}$ range \citep[see Figure 6 in ][]{Fujii2011}.  However the expectation for early O stars would be space velocities more in the region of 40 km~s$^{-1}$.  We can ask now whether this makes sense compared with our results on the diaspora of O Stars around Westerlund 2.  Given that we now have, from Table~\ref{tab:gaia}:

\begin{itemize}
\item Up to 7 O-star ejections in the neighbourhood, exhibiting extinctions and $K$ magnitudes compatible with those of O-type members of Wd2.  Radial velocities are presently unmeasured for 5 of them.  Objects \#1338 and \#1273 are in this group of 7, and have full space motions that fit in with 'runaway' designation, even at a distance of 
4 kpc, that we regard as at the low end of the likely distance range to Wd2 and associated objects.  The longest timescale since ejection is just over 1 Myr (for object \#1374).  For \#1374 and \#1046 the relative PMs are already so large, it goes against Occam's Razor not to regard them as most likely ejections from Wd2. There are thus 3 stars, \#1102, \#1133 and \#1236, with more modest relative PMs for which the case for association and ejection is not quite as complete.  It is noteworthy that \#1236 might be a partner to \#1273, lying directly on the opposite side of the cluster centre (see Figure~\ref{fig:pm-map}). 
\item A demonstration that WR 20c and WR 20aa have proper motions consistent with ejection from Wd2 also within the last million years.  
\item MSP 18 and MSP 171, the $\sim$O5 stars studied by \citet{Rauw2011} projecting onto the cluster core.  Their space motions away from Wd2 are respectively 29 and 37 km s$^{-1}$ and are dominated by the distance-independent radial velocities.  In view of their sky positions close to the centre of Wd2 this is as expected.   

\end{itemize}

\noindent
This adds up to 8 convincing ejections, and 3 more that could be, within the 1.5$\times$1.5 sq.deg. box examined.  In view of WR 21a's PM and relative PM being almost entirely in Galactic longitude and, so, potentially attributable to being more distant than Wd2, we do not include it in the tally.

The prediction of \citet{Fujii2011} for ejected numbers has been more than met. %A precise fix on this demands further spectroscopic follow up to measure more radial velocities.  %But, as things stand, \#1338 would be predicted to exhibit a proper motion of $\sim$1.2 mas yr$^{-1}$ at a distance of 5~kpc, if it was indeed ejected from Wd2 with a full space velocity of 40 km s$^{-1}$.  Based on predicted performance figures\footnote{see figures for DR2 at https://www.cosmos.esa.int/web/gaia/dr2 }, this should be measured by the Gaia mission to a precision of 5-10\%.  Object \#1273 is much more of a stretch to accommodate: its relative radial velocity of only $\sim$11 km s$^{-1}$ is no more than comparable with the likely cluster velocity dispersion \citep[see the supplementary data presented by ][]{Fujii2011}.  To be consistent with escape (at 40 km s$^{-1}$), this object's proper motion would need to be almost half as much again as the estimate for \#1338.
%Of course, it remains to be seen if the parallaxes of these two stars are both good matches to Wd2, to within the errors (5-6$\sigma$ detections are likely, Katz \& Brown 2017).
There is potential here for the ejected O-star population to be a large fraction of the total remaining in the cluster.  The cluster core only contains $\sim$30 O stars altogether \citep{Vargas2013}, implying an ejection efficiency of between 21 and 27\%.  But the data in Figure 3 of \cite{Fujii2011} would favour only 5 to 10\% .  There is a clear candidate 'bully binary' in this cluster in WR~20a, for which the estimated total mass is $\sim$165~M$\odot$ \citep{Bonanos2004}, but can it have promoted so many ejections, and why would there be a preferred plane for the process?  

A separate clue to what might be going on here comes from the structure of Wd2 itself: \citet{Zeidler2017} have commented on the presence of the 'northern clump' offset from the main cluster by almost an arcminute: they argue that this would be consistent with late merging behaviour if Wd2 has built up from sub-clusters.  These authors, along with \citet{Furukawa2009} and others studying the molecular gas, have also raised the possibility that a larger scale molecular cloud collision was the ultimate trigger for Wd2's formation.  If so, the modelling of \cite{Lucas2018} on the creation of an O-star halo, as a product of earlier molecular cloud/proto-cluster merging history, becomes directly relevant.  A feature of these models is that lower space velocities ($< 20$ km s$^{-1}$) are expected, and the distribution of objects within the halo should follow what would once have been tidal tails in the merging process.  Whilst this might explain the existence of the preferred proper motion plane -- which is roughly north-south -- higher measured speeds are more in keeping with dynamical ejection. Even at the near distance of $\sim 4$ kpc, the 4 complete space velocities now in hand scatter from 26 up to 40 km~s$^{-1}$ (see Table~\ref{tab:gaia}).  Interestingly, the follow-up simulations by \cite{Fujii2012} did examine the option of merging 4 sub-clusters as a model for Wd2: the key result apparent in Figure 7 of their paper is that the runaway fraction would then rise to 15 to 20\% (depending on where the stellar mass cut is made).  This is more promising.

Ultimately, it is certainly credible that both dynamical ejection and sub-cluster collision and merging can operate.  In the case of \#1273, it was pointed out in Section~\ref{sec:rv-implications} that its RV is similar to one of the molecular gas components plausibly just in front of Wd2. This may not be a coincidence.   What is clear, and has no dependence on the still uncertain distance to Wd2, is that the ejections identified so far have taken place over the last million years.  If this phase of activity was initiated by the merging of pre-existing sub-clusters, it leaves open a longer timescale for the WR stars to evolve within.  Indeed a search over a wider area for more ejected objects might provide evidence for a longer timescale.
%We may find there is an on-sky distinction, such that there is an inner ring of more slowly moving O stars that represent the detritus of past merging behaviour (e.g. \#1086), with the runaways resulting from 'harder' dynamical interaction taking over at larger displacements (\#1338, the clearest example here). 

A further point to make is that all the likely ejections are either earlier-type O or WR stars.  For 3 of them (\#1046, \#1102, \#1133) we do not have spectroscopic parameters to hand.  Among these, \#1046 has the coolest effective temperature estimate ($\sim35$ kK) of the set, based on its photometric properties.  On the \cite{MSH2005} scale, this corresponds to $\sim$O7 spectral type.  The status of \#1273 and \#1338 is now very clear: these are certainly O4 stars, with masses most likely $\sim$50~M$_{\odot}$ or more (Section~\ref{sec:parms}).  Better parameters from spectroscopy are clearly needed to e.g. tie down the higher stellar mass favouritism of the ejection process and its efficiency. 

In contrast, there remains a set of 5 mostly cooler O stars (\#0673, \#0693, \#0893, \#0994 and \#1164) from the upper half of Table~\ref{tab:list} that show little relative PM and no clear sign, so far, of associated clusters.  Only for \#1567 is there sign it has its own cluster.  As found in other work there remains scope for isolated formation and/or stochastic sampling of the IMF.  How far these stars are from Wd2 is another question awaiting follow up spectroscopy.

The object, WR 20c, remains a conundrum.  Its kinematics, as revealed by its proper motion, are clearly consistent with ejection from Wd2.  But how then are we to understand the huge differential extinction of around 4 visual magnitudes with respect to Wd2 -- combined with no sign in WISE mid-IR data of local dust warming.  This warrants further investigation, and we note that the astrometry on this object -- no doubt thanks to its high extinction -- is the least robust at the present time.

\section{Conclusions} 
\label{sec:conclusions}

New observations presented here have enabled the measurement of stellar parameters and radial velocities of two recently-identified massive stars ($M > 40M_{\odot}$) at projected distances of $\sim$15 and $\sim$30 pc from Wd2.  At the working distance of 5 kpc, the absolute magnitudes of VPHAS-OB1-01273 and VPHAS-OB1-01338 (and their RV comparison stars in Wd2) fit comfortably with the class V values for their spectral types set out by \cite{MSH2005}.  The radial velocity and proper motion data now in hand for both indicate they are moving away from this young massive cluster, and that they meet the standard runaway criterion, of a space velocity exceeding $\sim$25 km s$^{-1}$ or more (at a distance of at least 4 kpc).  Repeat observations indicate a low risk that the measured radial velocity of either object is significantly influenced by so-far undetected binarity.

The second main result, based on Gaia DR2 proper motions, concerns the frequency and on-sky distribution of recent early O and WR star ejections from Wd2 (located within a box of 1.5$\times$1.5 sq.deg.).  The number of likely ejections (between 8 and 11), their typical speeds, and their near N--S alignment cannot be explained by either sub-cluster merging or dynamical ejection alone.  A combination of both might work.  
%It seems likely that previous suggestions Wd2 is the product of sub-cluster merging \citep{Zeidler2017} or molecular cloud collision \citep{Furukawa2009} will prove relevant.  

%We note that the matter of the distance to Wd2 remains difficult.  We have used 5 kpc as a working distance, whilst aware that the median cluster proper motion derived from Gaia DR2 data points to 6--7 kpc.  This is not a problem for the kinematic results we have presented in that a larger distance raises the absolute space motions even more. 
%To complete the picture for these two instances, proper motions for both Wd2 cluster members in the mean and these two stars will be needed.  Since the $G$ magnitudes involved are 13--14, the proper motions delivered in Gaia Data Release 2 due in spring 2018 should be of good quality, even if the distance constraints from the accompanying parallaxes are not expected to provide decisive depth resolution (for distances expected to be in the range 4--6 kpc).

Spectroscopic follow up of a quality sufficient to provide more radial velocities to a precision of 2--3 km s$^{-1}$, and stellar mass estimates, for more of the recently uncovered O stars in the area is needed to work through what is proving to be an interesting and informative example of a massive-cluster O-star diaspora.  In discussing the results now available, we can already see that insights into the history of Wd2's formation may be forthcoming.  The work so far has neatly distinguished 'runaway' O stars ejected in the last million years, from those exhibiting little relative motion that remain as isolated field O stars, mostly with little sign of their own clusters.  A search over a wider sky area than here could be of interest to rule on whether Wd2 has been ejecting massive stars for longer than a million years.  Given the young age of Wd2, more than $\sim$2 million years is very unlikely.
%may betray a sub-cluster merging history \citep{Lucas2018}, while at larger angular separations dynamically-ejected runaways may take over, providing direct tests of existing quantitative models \citep[e.g.][]{Fujii2011}.  

%So far, we have only one example (from an initial selection of 19 O candidates) of an O star likely to be close to Wd2 that appears to be accompanied by a cluster containing, at least, B/A stars. Beyond this, there is no evidence available yet that demands in situ massive-star formation spread across 'greater Wd2'.  If these stars have formed away from the dense environment of the Wd2 cluster itself, the stochastic picture described by \citet{Parker2007} is relevant. 

In this study of Wd2 we began with the hypothesis that extinction can provide a first cut on identifying a cluster's more immediate associated population.  It is an easy cut to make, given the good quality extinction data available from MS-II, and it has proved useful.  Making this distinction helped draw attention to a separate, more reddened physical grouping, made up of a tight clustering of O stars at $\ell \simeq 283.88^{\circ}$, $b \simeq -0.91^{\circ}$, embedded in a lower density $\sim$20 arcmin halo.  This first impression is now backed up by the relative proper motion data.  All the new candidates for ejection from Wd2 have come from an initial selection of stars with similar reddening to Wd2.

Now that we can (i) find O and early B stars with ease across the Galactic Plane to many kiloparsecs and behind up to $\sim$10 magnitudes of optical extinction, (ii) combine these findings with Gaia proper motions, the path to a much richer understanding of the sites of their formation and subsequent kinematic histories is well and truly open.

%But for now we see extinction as providing a first-order sift of candidate objects, if applied conservatively to a well-constrained object sample.  Indeed, whilst 6-7 magnitudes of visual extinction continues as a nuisance at optical wavelengths, it is for certain no bar to a rewarding study of spatially well-resolved young Galactic environments. 

\section*{Acknowledgements}

This paper is based on data products from observations made with ESO Telescopes at the La Silla Paranal Observatory under programme ID 095.D-0843.  It also rests on results obtained directly from the VST Photometric Hα Survey of the Southern Galactic Plane and Bulge (VPHAS+, www.vphas.eu; ESO programme 177.D-3023). Use is also made of data products from the Two Micron All Sky Survey, which is a joint project of the University of Massachusetts and the Infrared Processing and Analysis Center/California Institute of Technology, funded by the National Aeronautics and Space Administration and the National Science Foundation. WISE and AAT service programme acknowledgements.  This work has made use of data from the European Space Agency mission Gaia (https://www.cosmos.esa.int/gaia), processed by the Gaia Data Processing and Analysis Consortium (DPAC, https://www.cosmos.esa.int/web/gaia/dpac/consortium). Funding for the DPAC has been provided by national institutions, in particular the institutions participating in the Gaia Multilateral Agreement.

This research made use of TopCat \citep{Taylor2005}.

% The authors wish to thank the referee for helpful comments that have improved this paper.

JED and MM acknowledge the support of a research grant funded by the Science, Technology and Facilities Council of the UK (STFC, ref. ST/M001008/1).  AH acknowledges support from Spanish MINECO under project grants AYA2015-68012-C2-1 and SEV2015-0548 and from the Gobierno de Canarias under project grant  ProID2017010115. NJW acknowledges receipt of an STFC Ernest Rutherford Fellowship (ref. ST/M005569/1).

We thank an anonymous referee for comments that have helped improve this paper's content.

\bibliographystyle{mnras}
\bibliography{JEDpapers}

\appendix

\section{Further positional information}

The table below supplements Table~\ref{tab:list} with celestial coordinates for every object and identifies the cross-matched 19-digit Gaia DR2 source name.    For completeness, the additional objects discussed in connection with Table~\ref{tab:gaia} are also included.  It can be seen that the typical cross-match distance is most often between 0.1 and 0.15 arcsecs.  In no case is there difficulty in identifying the counterpart source.

\begin{table*}
\caption{The columns contain: the full MS-II object name; object position as given by MS-II (RA, Dec J2000); the linked Gaia source identification number; cross-match distance in arcsec between the MS-II and Gaia DR2 positions, rounded to 2 significant figures. Note that the celestial coordinates are also rounded, rather than truncated, to two decimal places (in seconds of time and arc).  The order of listing in the table is as in Table 1, with the 3 already known objects added on in section (c).  Objects appearing in Table 7 are marked with an asterisk in the final column.}
{\centering
\begin{tabular}{lccccc}
\hline
MS-II name & \multicolumn{2}{c}{RA,Dec J2000} & Gaia Source \# & offset & \\
           &    (hh:mm:ss) & (dd:mm:ss)       &                & (arcsec) & \\  
\hline             
(a)    & & & & \\
\hline
VPHAS-OB1-00673 & 10:18:22.62 & -57:31:08.63 & 5258689821862198272 &  0.08 & \\ 
VPHAS-OB1-00693 & 10:19:01.47 & -57:26:34.40 & 5258710854331053696 &  0.12 & \\
VPHAS-OB1-00708 & 10:16:53.91 & -57:55:02.11 & 5258584101263763200 &  0.09 & \\
VPHAS-OB1-00893 & 10:19:47.82 & -57:50:38.64 & 5258666800837002880 &  0.19 & \\
VPHAS-OB1-00986 & 10:19:53.04 & -58:00:00.42 & 5258659074204741760 &  0.16 & \\
VPHAS-OB1-00987 & 10:19:33.60 & -58:04:05.07 & 5258656909541094400 &  0.15 & \\ 
VPHAS-OB1-00994 & 10:21:20.56 & -57:43:09.40 & 5255690873859259904 &  0.14 & \\ 
VPHAS-OB1-01046 & 10:25:35.77 & -57:00:00.07 & 5351803514564210560 &  0.12 & * \\
VPHAS-OB1-01102 & 10:23:46.52 & -57:34:15.51 & 5351760152573602944 &  0.12 & * \\
VPHAS-OB1-01133 & 10:24:19.34 & -57:33:02.48 & 5351760908487879296 &  0.12 & * \\
VPHAS-OB1-01164 & 10:22:18.41 & -58:02:16.66 & 5255647507548038272 &  0.10 & \\
VPHAS-OB1-01236 & 10:24:43.47 & -57:37:15.94 & 5351757438154160000 &  0.13 & * \\ 
VPHAS-OB1-01273 & 10:23:26.43 & -57:56:03.67 & 5255669399023171456 &  0.16 & * \\
VPHAS-OB1-01308 & 10:23:23.50 & -58:00:20.80 & 5255667681036173568 &  0.13 & * \\
VPHAS-OB1-01338 & 10:26:03.10 & -57:31:43.06 & 5351717851422618496 &  0.06 & * \\
VPHAS-OB1-01356 & 10:23:59.03 & -58:00:48.44 & 5255668024633629824 &  0.12 & \\
VPHAS-OB1-01374 & 10:22:32.65 & -58:20:31.99 & 5255633871052073728 &  0.14 & * \\
VPHAS-OB1-01550 & 10:25:41.75 & -58:05:52.80 & 5255622016940658304 &  0.18 & \\ 
VPHAS-OB1-01567 & 10:24:42.37 & -58:20:31.94 & 5255590264223148032 &  0.11 & \\
\hline 
(b) & & & & \\
\hline
VPHAS-OB1-00685 & 10:17:10.97 & -57:47:57.31 & 5258678689306546304 &  0.11 & \\
VPHAS-OB1-00785 & 10:23:19.04 & -56:48:47.93 & 5354818130613185792 &  0.12 & \\
VPHAS-OB1-00826 & 10:19:29.49 & -57:43:37.95 & 5258671959106226432 &  0.13 & \\
VPHAS-OB1-00879 & 10:19:34.95 & -57:51:44.54 & 5258666564627442304 &  0.17 & \\
VPHAS-OB1-00881 & 10:20:09.94 & -57:44:39.35 & 5258669038528673920 &  0.16 & \\
VPHAS-OB1-00896 & 10:19:14.68 & -57:57:59.27 & 5258659864478620800 &  0.13 & \\
VPHAS-OB1-00904 & 10:18:57.74 & -58:02:31.51 & 5258657837254006400 &  0.12 & \\
VPHAS-OB1-00918 & 10:19:06.93 & -58:02:14.03 & 5258658180851400320 &  0.13 & \\
VPHAS-OB1-00919 & 10:19:08.56 & -58:01:55.61 & 5258658176542224256 &  0.13 & \\
VPHAS-OB1-00921 & 10:19:09.69 & -58:01:59.30 & 5258657974692975744 &  0.12 & \\
VPHAS-OB1-00925 & 10:19:09.18 & -58:02:26.35 & 5258657974680705024 &  0.09 & \\
VPHAS-OB1-00930 & 10:19:14.02 & -58:02:08.06 & 5258658009052717568 &  0.15 & \\
VPHAS-OB1-00931 & 10:19:12.79 & -58:02:23.10 & 5258657974692975872 &  0.15 & \\
VPHAS-OB1-00934 & 10:19:01.03 & -58:04:49.65 & 5258657424937135232 &  0.14 & \\
VPHAS-OB1-00938 & 10:20:40.41 & -57:45:02.26 & 5258691475439095424 &  0.15 & \\
VPHAS-OB1-00953 & 10:24:29.87 & -56:58:08.61 & 5351810901907785088 &  0.09 & \\
VPHAS-OB1-00958 & 10:19:23.76 & -58:02:42.80 & 5258657184419001472 &  0.15 & \\
VPHAS-OB1-00965 & 10:19:51.48 & -57:58:02.22 & 5258659417802053888 &  0.15 & \\	
VPHAS-OB1-00968 & 10:19:12.15 & -58:06:04.33 & 5258656634663154048 &  0.11 & \\
VPHAS-OB1-01086 & 10:24:06.02 & -57:27:34.29 & 5351762489035515904 &  0.10 & \\
VPHAS-OB1-01119 & 10:25:02.61 & -57:21:47.33 & 5351766715283439104 &  0.08 & * \\
\hline
(c) & & & & \\
\hline
MSP 18 & 10:24:02.44 & -57:44:36.05 &  5255678500030907904 &  0.11 & * \\
MSP 171 & 10:24:04.90 & -57:45:28:35 & 5255678126396953344 &  0.11 & * \\
WR 21a & -- & --  &  5351703390282380800 & -- & * \\ 
\hline
\end{tabular}
}
\label{tab:crossref}
\end{table*}

\label{lastpage}

\end{document}